\documentclass[12pt,iop,numberedappendix]{emulateapj}
\usepackage{epsfig,graphicx}

\newcommand{\etal}{et al.}
\newcommand{\halpha}{\hbox{{$\rm H\alpha$}}}
\newcommand{\hii}{\hbox{\ion{H}{2}}}
\newcommand{\nii}{\hbox{[\ion{N}{2}] }}
\newcommand{\sii}{\hbox{[\ion{S}{2}] }}
\newcommand{\hone}{\hbox{{\rm H{\small I}}}}
\newcommand{\htwo}{\hbox{${\rm H_2}$}}
\newcommand{\cone}{\hbox{${\rm CO \ {\it J}=1-0}$}}
\newcommand{\ctwo}{\hbox{${\rm CO \ {\it J}=2-1}$}}

\newcommand{\nmol}{\hbox{$\rm N_{mol}$}}
\newcommand{\cf}{\hbox{$\rm X_{CO}$}}
\newcommand{\um}{\hbox{$\mu$m}}

\newcommand{\Shtwo}{\hbox{$\Sigma_{\rm H_2}$}}
\newcommand{\Ssfr}{\hbox{$\Sigma_{\rm SFR}$}}
\newcommand{\Sgas}{\hbox{$\Sigma_{\rm gas}$}}
\newcommand{\Sfuv}{\hbox{$\Sigma_{{\rm FUV}+24\,\um}$}}
\newcommand{\Shal}{\hbox{$\Sigma_{{\rm H\alpha}}$}}
\newcommand{\Scom}{\hbox{$\Sigma_{{\rm H\alpha}+24\,\um}$}}
\newcommand{\Smir}{\hbox{$\Sigma_{24\,\um}$}}
\newcommand{\SFRfuv}{\hbox{${{\rm FUV}+24\,\um}$}}

\newcommand{\SFRcom}{\hbox{${\rm H\alpha}+24\,\um$}}

\newcommand{\Jkmsec}{\hbox{$\rm Jy \ km \ sec^{-1}$}}
\newcommand{\JBkmsec}{\hbox{$\rm Jy~Beam^{-1} \ km \ sec^{-1}$}}

\newcommand{\JB}{\hbox{\rm Jy~Beam$^{-1}$ }}
\newcommand{\mJB}{\hbox{\rm mJy~Beam$^{-1}$ }}
\newcommand{\kmsec}{\mbox{km \ sec$^{-1}$}}

\newcommand{\sfe}{\mbox{$\rm SFE$}}
\newcommand{\sigsct}{\mbox{$\rm \sigma_{i}$}}
\newcommand{\sigsctfit}{\mbox{$\rm {\bar \sigma}_{i,fit}$}}
\newcommand{\sigfit}{\mbox{$\rm \sigma_{mol,fit}$}}
\newcommand{\nmolfit}{\mbox{$\rm {\bar{N}_{mol, fit}}$}}
\newcommand{\msun}{\hbox{$\rm M_{\odot} $ }}
\newcommand{\msunpc}{\hbox{$\rm M_{\odot} \ pc^{-2}$}} 
\newcommand{\msunyr}{\mbox{$\rm M_{\odot} \ yr^{-1}$}}
\newcommand{\msunyrpc}{\mbox{$\rm M_{\odot} \ Gyr^{-1} \ pc^{-2}$}}

\newcommand{\fde}{\hbox{${f_{\rm DE}}$}}
\newcommand{\ffuv}{\hbox{${f_{\rm FUV}}$}}
\newcommand{\fdig}{\hbox{${f_{\rm DIG}}$}}
\newcommand{\fmir}{\hbox{${f_{\rm MIR}}$}}

\newcommand{\imhal}{\hbox{$\rm I_{H\alpha}$}}
\newcommand{\immir}{\hbox{$\rm I_{MIR}$}}
\newcommand{\erfuv}{\hbox{$\rm \sigma_{FUV}$}}
\newcommand{\erhal}{\hbox{$\rm \sigma_{H\alpha}$}}
\newcommand{\ermir}{\hbox{$\rm \sigma_{MIR}$}}
\newcommand{\optrad}{\hbox{$\rm R_{25}$}}
\newcommand{\tdep}{\hbox{$\tau_{dep}$}}

\shorttitle{Molecular Gas Star Formation Law in NGC~4254}
\shortauthors{Rahman et al.}

\begin{document}


\title{CARMA Survey Toward Infrared-bright Nearby Galaxies (STING): 
Molecular Gas Star Formation Law in NGC~4254}

\author{
Nurur Rahman\altaffilmark{1}, 
Alberto D. Bolatto\altaffilmark{1},
Tony Wong\altaffilmark{2}, 
Adam K. Leroy\altaffilmark{3}, 
Fabian Walter \altaffilmark{4},
Erik Rosolowsky\altaffilmark{5}, 
Andrew A. West\altaffilmark{6}, 
Frank Bigiel \altaffilmark{7}, 
J\"{u}rgen Ott\altaffilmark{8}, 
Rui Xue\altaffilmark{2}, 
Rodrigo Herrera-Camus\altaffilmark{1},
Katherine Jameson\altaffilmark{1}, 
Leo Blitz\altaffilmark{7}, 
Stuart N. Vogel\altaffilmark{1}
}

\altaffiltext{1}{Department of Astronomy, University of Maryland, 
College Park, MD 20742, USA; nurur@astro.umd.edu}
\altaffiltext{2}{Department of Astronomy, University of Illinois, 
Urbana-Champaign, IL 61801, USA}
\altaffiltext{3}{National Radio Astronomy Observatory, Charlottesville, 
VA , USA}
\altaffiltext{4}{Max-Planck-Institute fur Astronomie, Konigstuhl 17, 
69117, Heidelberg, Germany}
\altaffiltext{5}{I. K. Barber School of the Arts \& Science, University 
of British-Columbia, Kelowna, BC V1V1V7, Canada}
\altaffiltext{6}{Department of Astronomy, Boston University, Boston, 
MA 02215, USA}
\altaffiltext{7}{Department of Astronomy, University of California, 
Berkeley, CA 94720, USA}
\altaffiltext{8}{National Radio Astronomy Observatory, Socorro, NM 87801, 
USA}

\begin{abstract}
This study explores the effects of different assumptions and
systematics on the determination of the local, spatially resolved 
star formation law.  Using four star formation rate (SFR) tracers
(\halpha\ with azimuthally averaged extinction correction,
mid-infrared 24 \um, combined \halpha\ and mid-infrared 24 \um, and
combined far-ultraviolet and mid-infrared 24 \um), several fitting
procedures, and different sampling strategies we probe the relation
between SFR and molecular gas at various spatial resolutions (500 pc
and larger) and surface densities ($\Shtwo\approx10-245$ \msunpc)
within the central $\sim6.5$ kpc in the disk of NGC~4254. 
We explore the effect of diffuse emission using an unsharp masking 
technique with varying kernel size. The fraction of diffuse emission, 
\fde, thus determined is a strong inverse function of the 
size of the filtering kernel.
We find that in the high surface brightness regions of NGC~4254 the 
form of the molecular gas star formation law is robustly determined 
and approximately linear ($\sim 0.8-1.1$) and independent of the 
assumed fraction of diffuse emission and the SFR tracer employed. 
When the low surface brightness regions are included, the slope of 
the star formation law depends primarily on the assumed fraction of 
diffuse emission. In such case, results range from linear when the 
fraction of diffuse emission in the SFR tracer is $\fde\lesssim30\%$ 
(or when diffuse emission is removed in both the star formation and 
the molecular gas tracer), to super-linear ($\sim 1.4$) when
$\fde\gtrsim50\%$.  We find that the tightness of the correlation
between gas and star formation varies with the choice of star
formation tracer. The 24 \um\ SFR tracer by itself shows the tightest
correlation with the molecular gas surface density, whereas the
\halpha\ corrected for extinction using an azimuthally-averaged
correction shows the highest dispersion.  We find that for $\rm
R<0.5\optrad$ the local star formation efficiency is constant and
similar to that observed in other large spirals, with a molecular 
gas depletion time $\tdep\sim2$~Gyr.
\end{abstract}


\section{Introduction}
The formation and evolution of galaxies is driven by the complex
processes of star formation (SF) that occur inside them. Some 
galaxies produce stars at very low rates $\lesssim0.1$ \msunyr, 
some do at modest rates $\sim1$ \msunyr, while others host ongoing 
starbursts with SFR, $\sim10-1000$ \msunyr.  
The processes responsible for converting gas into stars in various
galactic environments are still poorly understood. Observations 
find that the SFR and the gas content in galaxies are related by,
\begin{equation}
\rm \Sigma_{\rm SFR} = A \ \Sigma_{\rm gas}^N, 
\label{sfl}
\end{equation}
where \Ssfr\ and \Sgas\ are the star formation rate surface density
and the gas (atomic and molecular) surface density, respectively; and
$\rm A$ is the normalization constant representing the efficiency of 
the processes (Schmidt 1959, 1963; Sanduleak 1969; Hartwick 1971;
Kennicutt 1989). For disk averaged surface densities, both normal
star-forming and starburst galaxies follow this relationship with a
power law index of $\rm N \sim1.3-1.5$ for total gas (Kennicutt 1989,
1998a,b). This relationship between the gas and SFR surface densities 
is commonly referred to as the Schmidt-Kennicutt SF law. This is in 
principle consistent with large scale gravitational instability being 
the major driver (Quirk 1972; Madore 1977).

Although spatially unresolved studies of \hone, CO, and SFR are useful 
for characterizing global disk properties, understanding the mechanisms
behind the SF law requires resolved measurements. Only recently has 
it become possible to probe the form of the gas-SF relationship on 
kpc and sub-kpc scales, through the availability of high-resolution
interferometric \hone\ and single-dish CO observations and of a suite 
of multi-wavelength SFR tracers. Studies of the local SF law on nearby
galaxies provide substantial evidence that the molecular gas is
well-correlated with the SFR tracers, whereas the atomic gas shows
little or no correlation with SF activity (e.g., Wong \& Blitz 2002; 
Bigiel \etal. 2008).
This is a natural consequence of stars forming out of giant molecular
clouds (GMCs), as we observe in the local universe. Moreover, it has
long been known that the spatial distribution of CO emission follows
closely that of the stellar light and \halpha\ (Young \& Scoville
1982; Scoville \& Young 1983; Solomon \etal\ 1983; Lord \& Young 1990;
Tacconi \& Young 1990; Boselli \etal\ 1995).
 
Spatially resolved SF law studies frequently reach dissimilar
conclusions on the value of the exponent in Equation \ref{sfl} when 
relating {\em molecular gas} to SFR. Hereafter we will express the 
exponent as \nmol\ to represent the molecular gas SF law.
Wong \& Blitz (2002) used azimuthally averaged radial profiles for gas
and SFR in a sample of seven molecule-rich spiral galaxies, finding
that the best fit power law index for the molecular gas and SFR
density radial profiles is $\nmol\sim0.8 - 1.4$, very dependent on the
extinction correction applied to their SFR tracer (\halpha).  Boissier
\etal\ (2003) used CO observations along the major axes of sixteen
disk galaxies with spatial resolution of $\sim1-4$~kpc to carry out 
a similar study. They found a somewhat steeper exponent $\rm
\nmol\sim1-2.3$ with respect to the molecular gas, but with a
different choice for the extinction correction and different fitting
methodologies.  
Sampling various spatial scales and surface densities of a sample 
of twenty three disk galaxies Komugi \etal\ (2005) found $\nmol \sim1.3$ 
using extinction corrected \halpha\ as the SFR tracer.
In the most recent comprehensive study, Bigiel \etal\
(2008) analyzed a sample of 18 normal disk and irregular galaxies
using a combination of {\em GALEX} far-ultraviolet (FUV, 1350-1750
\AA) emission and {\em Spitzer} 24 \um\ to trace SFR, and CO $2-1$ 
for the molecular gas at a spatial resolution $\sim750$~pc. They found 
a best fit slope $\rm \nmol \approx1.0$ for the power-law relation
between \Shtwo\ and \Ssfr.

This spread in the value of the power law index is observed by in-depth 
case studies of just one galaxy. An example is M~51, where Kennicutt 
\etal\ (2007) studied the relation between gas and SFR on $\sim0.3-1.8$ 
kpc scales sampling the distribution of emission in circular apertures 
centered on \hii\ regions to find $\nmol\sim1.37-1.56$, depending on 
the spatial scale considered. 
Another example is NGC~7331 studied by Thilker \etal\ (2007). The 
authors find $\nmol\sim1.64$ using bolometric (combining ultraviolet 
and infrared luminosity) SFR tracer at 400 pc resolution. By comparison, 
Schuster \etal\ (2007) used the $\lambda=20$~cm radio-continuum as SFR 
tracer and \ctwo\ data to find a much shallower molecular gas power-law
$\nmol\lesssim1$, changing with galactocentric distance. Similarly,
Blanc \etal\ (2009) studied the central $\sim$4 kpc of M~51 using
optical spectroscopic \halpha\ data at 170 pc resolution, finding a
slightly sub-linear relationship ($\nmol \sim 0.82 \pm 0.05$).
A very recent example is M~33, where Verley \etal\ (2010) 
employed a range of methods of data sampling, fitting techniques, and 
SFR tracers to find that the functional form of molecular gas-SFR 
relation varies ($\nmol \sim 1.2-1.8$) depending on the choice of SFR 
tracers, data sampling and fitting techniques. For the same galaxy but 
with a different SFR tracer Heyer \etal\ (2004) found $\nmol\sim1.4$.

The spread in the value of the power-law index within and among
galaxies may be intrinsic and contain valuable astrophysical
information, or be entirely attributable to the different choices of
gas and SFR tracers, methodologies for internal extinction correction,
differences in the CO-to-H$_2$ conversion, or the range of spatial
scales probed.  It is important to keep in mind that the choice of SFR
tracers and spatial scales means that different studies effectively
sample different time scales, thus the SF history of any particular
galaxy potentially plays an important role in determining the result
of the measurement. It is also possible that these differences
correspond to a spectrum of physical SF mechanisms present in a wide
range of environments: in that case, the local SF law would not be
universal. It is, therefore, vital to understand the impact of
systematics on the measurement of the parameters of the local SF law.
Whether the local SF law is linear or non-linear has implications for
the dominant SF mechanisms as well as for modeling efforts.

The objective of this paper is to explore the molecular gas SF law in
the galaxy NGC~4254 (M~99), at $\sim$500~pc and $\sim1$ kpc scales
using different SFR tracers. This galaxy has been the subject of a
number previous SF law studies (Kennicutt 1989; Boissier \etal\ 2003;
Komugi \etal\ 2005; Wilson \etal\ 2009), although not at such high
resolution.  This is a pilot study using observations obtained by the
Survey Towards Nearby Infrared-bright Galaxies (STING; Bolatto \etal\
in prep.), which employs the Combined Array for Research in Millimeter 
Astronomy (CARMA) interferometer.

In this paper we investigate the impact of various methodological
aspects related to local SF law study. Our study focuses on 1) the 
use of different SFR tracers and the scatter associated with those
tracers, 2) the role of the diffuse emission, a component of the
integrated disk emission which is not necessarily related to the 
star-forming regions, and 3) the role of fitting methodologies and 
data sampling strategies in determining the functional form of the SF
law. We do not explore the role of variations in the stellar initial 
mass function (IMF), the CO-to-H$_2$ conversion factor, the
extinction correction, and various other assumptions pertinent to
local SF law studies. These issues will be addressed in the future 
using other sample galaxies of the STING survey.

The organization of the paper is as follows. In $\S$\ref{data} we
present the multi-wavelength data set, including a brief description
of the data products. A discussion of the sky background and the
extended diffuse emission (DE) is presented in $\S$\ref{back}. Section
$\S$\ref{result} contains our main results and general discussions. We
compare our results with the most recent studies of local SF law in
$\S$\ref{compare}. The main findings of our study are summarized in
$\S$\ref{summary}. A brief introduction to NGC~4254, details on the
construction of various data products such as surface density maps,
the discussion on the treatment of DE, and the details of the sampling
and the regression analysis can be found in several sections of the
Appendix.

\section{Data} 
\label{data}
The target of this study, NGC~4254, is an almost face-on
($i\sim30^{\circ}$) SA(s)c spiral located at a kinematic distance of
16.6 Mpc (Prescott \etal\ 2007). It lies in the outskirts of Virgo
cluster, $\sim1.2$ Mpc to the north-west from the cluster center in
projected distance. For our adopted distance to NGC~4254, 1\arcsec\
corresponds to $\approx80$ pc. The optical radius of this galaxy is
$\optrad\approx12.1$~kpc. See appendix \ref{appen:ngc4254} for more 
information on this galaxy.

We construct the molecular gas density maps using a combination of 
\cone\ emission obtained by CARMA interferometer for the STING\footnote
{http://www.astro.umd.edu/$\sim$bolatto/STING/} survey, and single-dish 
\ctwo\ observations obtained by the Institut de Radio Astronomie
Millimetrique (IRAM) 30~m telescope at Pico Veleta, Spain. 
These data are part of the HERA CO Line Extragalactic Survey (HERACLES) 
and were observed and reduced in the same manner as the first part of 
the survey described in Leroy et al. (2009). The full survey will be 
presented by (Leroy et al. in prep.).
To construct SFR maps of NGC~4254 we use ultraviolet (UV) images from
{\em GALEX} Nearby Galaxy Survey (NGS; Gil de Paz \etal\ 2007), and
H${\alpha}$ and 24 \um\ images from the {\em Spitzer} Infrared Nearby
Galaxies Survey (SINGS\footnote
{http://ssc.spitzer.caltech.edu/legacy/singshistory.html}; Kennicutt 
\etal\ 2003) archive. 

Here we describe the multi-wavelength data. The basic information for 
the data set is provided in Table \ref{image_table}. For various other 
properties of NGC~4254 the reader is referred to Table 1 of Kantharia 
\etal\ (2008).

\subsection{CARMA STING Data}
\label{stingdata}

The interferometric \cone\ map of NGC 4254 was obtained as part of the
STING survey using CARMA interferometer. The STING sample is composed
of northern ($\delta > -20^{\circ}$), moderately inclined ($i <
75^{\circ}$) galaxies within 45 Mpc culled from the {\it IRAS} Revised
Bright Galaxy Survey (RBGS; Sanders \etal\ 2003). These galaxies have
been carefully selected to have uniform coverage in stellar mass, SF
activities, and morphological types. The survey is complementary to
BIMA SONG (Helfer \etal\ 2003) but targeted to have better disk
coverage, sensitivity and resolution. The details of the CARMA STING
survey will be published in a forthcoming paper (Bolatto \etal, in
preparation).

Observations with the CARMA interferometer were conducted in the D
array configuration during June 2008 for a total of 8.5
hours. Passband and phase calibration were performed using 3C273, and
3C274 was used as a secondary phase calibrator to assess the quality
of the phase transfer and coherence. The absolute flux scale for the
interferometer was determined by observing Mars.  At $\lambda = 2.6$
mm the 6~m (10~m) diameter CARMA antennas have a primary beam FWHM of
90\arcsec\ (54\arcsec)which defines their effective field of view. The
observations were obtained using a Nyquist-sampled 19-pointing mosaic
pattern that provides an effective field of view of 100\arcsec\ in
diameter.  The synthesized beam produced using natural weighting has
4.3\arcsec\ FWHM, which is the resolution of the resulting map.

\begin{figure*}[t]
\epsscale{0.90}
\plotone{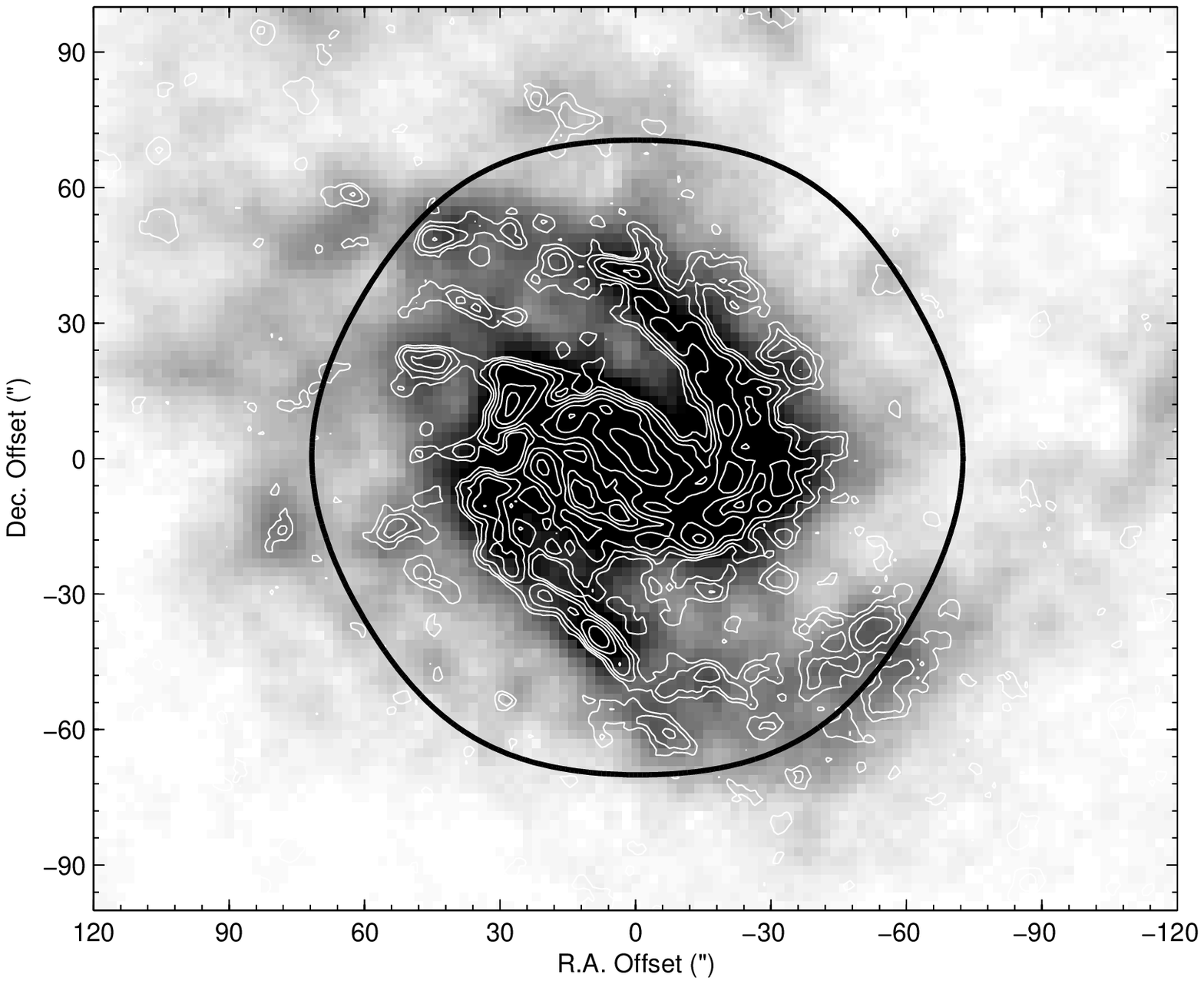}
\caption{The contours of velocity integrated CARMA \cone\ map overlaid 
on IRAM 30m single dish \ctwo\ map. The circle shows that part of the 
disk where CARMA interferometer observation is the most sensitive. 
Within this region the agreement between the two maps is excellent. 
The contours are in logarithmic scale with the levels at 2, 2.9, 4.3, 
6.3, 9.3, 13.6 and 20 \JBkmsec. The peak flux of CARMA map is 26 \JBkmsec. 
\label{comap}}
\end{figure*}

In our frequency setup, the receivers were tuned to the \cone\
transition at a rest frequency of 115.2712 GHz ($\lambda =2.6$ mm) in
the upper sideband, with the COBRA correlator configured to have three
spectral windows in each sideband, one in the low resolution 500 MHz
(16 channels) mode, and two partially overlapping in the 62 MHz (63
channels) mode. These were placed side by side in frequency with an
overlap of 10 MHz (6 channels). The resulting velocity range covered
was 290 km s$^{-1}$ with a intrinsic velocity resolution of 2.6 km
s$^{-1}$.  The maps were produced with 10 \kmsec\ velocity resolution.
The sensitivity of the interferometric map before combination with 
the single-dish data is $\sigma\approx22$ \mJB\ in 10 \kmsec\ wide
channels (see Fig. \ref{comap}). 

The high angular resolution measurements obtained with 
interferometers filter out the large spatial scales of the source, 
giving rise to the ``missing flux'' problem.  As a result, surface 
densities for extended sources derived from interferometric data 
alone may be underestimated. To overcome this limitation it is 
necessary to merge single-dish observation of the object with the 
interferometric data. 

The single dish IRAM \ctwo\ observations at $\lambda=1.3$ mm of 
NGC~4254 have $\sim12.5\arcsec$ beam and extend beyond the region 
mapped by STING. The sensitivity of the single dish map is 
$\sigma\approx27$ mK in 2.6 \kmsec\ wide channels. We have used a 
gain factor of $\rm T/S \sim 0.14 \ K/Jy$ appropriate for the IRAM 
30m single dish telescope and converted the units of the \ctwo\ 
data cube from Kelvin to \JB. 

Comparison of the enclosed fluxes shows that (assuming thermalized 
optically-thick CO emission, see below) the STING map recovers most 
of the single-dish flux in its inner 60\arcsec, progressively losing 
flux beyond that point (see Fig. \ref{comap}). We converted the \ctwo\ 
single dish cube to the equivalent \cone\ flux by applying a 
multiplicative factor of 4, following the assumption that the (peak) 
brightness temperature ($\rm T_{mb}$) is approximately the same for 
the \cone\ and \ctwo\ transitions. Before combination the CARMA cube 
was de-convolved using the implementation of the CLEAN algorithm in 
the MIRIAD task ``mossdi''. We have re-binned the velocity channels in 
the IRAM data cube to attain that of the CARMA cube. The combination 
between the CARMA and the IRAM cube was accomplished in the image 
plane (e.g., Stanimirovi\'c \etal\ 1999), using the MIRIAD task 
``immerge''. The spatial resolution of the combined map is the same 
as the CARMA \cone\ map. We have implicitly assumed a uniform filling 
factor while combining the two CO maps. 

Validation for the assumption of optically-thick thermalized CO 
emission comes from the comparison of single-dish fluxes. With a 
beam size of 45\arcsec, the global CO flux for NGC~4254 estimated 
in the FCRAO \cone\ survey is $(3000 \pm 850)$ \Jkmsec\ (Young 
\etal\ 1995). Our estimated flux using the IRAM \ctwo\ observations 
is $\sim2760$ \Jkmsec\ out to \optrad, showing that IRAM reproduces 
the FCRAO measurement within $\sim10\%$. This similarity shows that 
the assumption of identical brightness temperature for the $1-0$ 
and $2-1$ transitions is reasonable. 

Although the details of the study we present in the following sections 
depend somewhat on the combination of the interferometer with the single 
dish data, the major results are very much independent.
 
\tabletypesize{\scriptsize}
\begin{deluxetable*}{cccccc}[h!]
\tablewidth{0pt}
\setlength{\tabcolsep}{0.15 in} 
\tablecolumns{6}
\tablecaption{Basic Information of the Data Set}
\tablehead
{\colhead{Telescope}     
&\colhead{Wavelength} 
&\colhead{Pixel}  
&\colhead{FWHM}  
&\colhead{Sensitivity (1$\sigma$)}  
&\colhead{Sensitivity Unit}  
}
\startdata
{GALEX}      &0.2271 \um\   &1.5    &5.6       &$3.4\pm1.00$  
&$10^{-15}$ erg sec$^{-1}$ cm$^{-2}$ \\
KPNO             &0.6563 \um\   &0.3    &1.5   &$(6.7\pm1.5) \times 10^{-2}$ 
&$10^{-15}$ erg sec$^{-1}$ cm$^{-2}$ \\
{Spitzer}    &24     \um\   &1.5    &6.0       &$3.3\pm0.82$  
&$10^{-15}$ erg sec$^{-1}$ cm$^{-2}$ \\
CARMA        &2.6    mm     &1.0    &4.3   &0.22     &\JBkmsec   \\ 
IRAM         &1.3    mm     &2.0    &12.5  &1.00     &\JBkmsec   \\ 
$\Ssfr$      &--            &3.0    &6.0   &0.10     &\msunyrpc  \\ 
$\Shtwo$     &--            &3.0    &6.0   &3.70     &\msunpc
\enddata 
\tablecomments{\small 
The pixel resolution and the $\rm FWHM$ of FUV, H$\alpha$, and MIR 
24 \um\ maps are in unit of arcsec. The limiting sensitivities of 
the CARMA \cone\  and the IRAM \ctwo\ observations have been estimated 
from a single velocity channel map and 4 consecutive velocity channel 
maps, respectively, using $\rm \sigma = [\sigma_{ch} \sqrt{N}] \ \Delta v$, 
where $\rm \sigma_{ch}$ is the rms noise in a velocity channel map, $\rm N$ 
is the number of channel, and $\rm \Delta v$ is the velocity resolution. 
The velocity resolutions for the CARMA and the IRAM observations are 
$\rm \Delta v=10$ and 2.6 \kmsec, respectively. In this table only the 
surface densities are inclination corrected.} 
\label{image_table}
\end{deluxetable*}

\subsection{UV, Optical and Mid-infrared Data}

NGC~4254 has been observed in the near-ultraviolet (NUV, 1750-2750
\AA) by {\em GALEX}. To convert the map from NUV to FUV we use a
morphology dependent color correction $\rm (FUV-NUV)\approx0.46$,
following Gil de Paz \etal\ (2007). 
The morphology parameter $\rm T=5$ of NGC~4254 is obtained from the 
Third Reference Catalog (RC3; de Vaucouleurs et al. 1991). 
To correct the FUV map for line-of-sight Galactic extinction we use 
$\rm A_{FUV} \sim 8.24 \ E(B-V)$ (Weyder \etal\ 2007), where the 
Galactic reddening $\rm E(B-V) \approx0.04$ is estimated from Schlegel 
\etal\ (1998). The FUV map is converted to AB magnitudes using the 
following formula (Gil de Paz, private communication),
\begin{equation}
\rm m_{AB} = -2.5 \ \log[counts/sec] + 18.82
\end{equation}

The SINGS project public data archive provides calibrated and stellar
continuum subtracted \halpha\ image of this galaxy.  Comparing the
R-band and \halpha\ images we identify foreground stars which are then
masked, particularly those within the optical diameter. The resulting
image is then corrected for \nii\ $\lambda \lambda$6548, 6583
forbidden line emission and the transmission curve of the \halpha\
filter, using the factors obtained for this galaxy by Prescott \etal\
(2007).

The SINGS archive also provides the mid-infrared (MIR) 24 \um\ image, 
which s a scan map taken with the MIPS instrument on board the {\em
Spitzer} Space Telescope (Rieke \etal\ 2004). The MIPS data were
processed using the MIPS Instrument Team Data Analysis Tool (Gordon
\etal\ 2005). No stellar masking was necessary for the MIR map of
NGC~4254. Basic information of these images are given in Table 
\ref{image_table}.

\subsection{Data Products}
\label{product}

The spatial resolution of our study is limited by the point spread
function (PSF) of the MIR data, which has a FWHM of 6\arcsec. This 
angular scale corresponds to a physical length of $\sim480$ pc in 
the disk of NGC~4254. We should note that, although we approximate 
it as a Gaussian, the mid-infrared PSF is complex. It has prominent 
first and second Airy rings, with the second ring stretching out 
to $\sim 20\arcsec$. Nevertheless, approximately 85\% of the total 
source flux is contained within the central peak with FWHM of 
6\arcsec\ (Engelbracht \etal\ 2007).

The higher resolution \halpha\ and CO images were Gaussian-convolved
to have the same resolution and sampling as the MIR image. In both
cases, the convolution and regridding used the AIPS package\footnote{The 
Astronomical Image Processing System (AIPS) has been developed by the 
National Radio Astronomy Observatory (NRAO)}. No convolution was 
necessary for the FUV image, since it has a resolution similar to the 
MIR (see Table 1). For our high-resolution analysis we regrid the 
images to 3\arcsec\ pixels to Nyquist-sample the PSF.

\begin{figure*}[t]
\epsscale{1.1}
\plotone{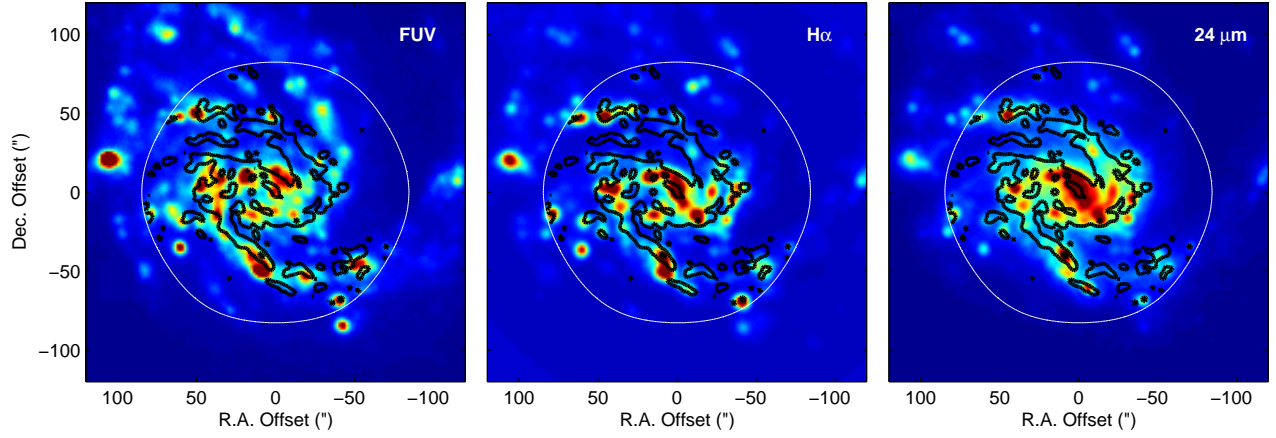}
\caption{Multi-wavelength maps of NGC~4254. The panels show FUV, 
\halpha, and MIR 24 \um\ emission maps from left to right. The images 
are shown in logarithmic scale. The black contours represent the 
layout of \cone\ map used in this study. The contours are in linear 
scale with the levels at 3 and 12 \JBkmsec. The figure shows a striking 
similarity between the distributions of hot dust traced by the MIR 
map and the cold molecular gas traced by \cone\ in both upper an 
lower arm of the galaxy. 
\label{sfmap}} 
\end{figure*}

Figure \ref{sfmap} shows the FUV, \halpha, and 24 \um\ images of
NGC~4254 used to construct the SFR maps in logarithmic color scale. 
The black contours correspond to the 3 and 12 \JB\ levels of the
\cone\ map. Figure \ref{sfmap} shows the striking similarity between 
the distributions of hot dust ($\rm T \sim100$~K) traced by the 24 
\um\ emission and the cold molecular gas traced by \cone\ in NGC~4254. 
Interestingly, the \cone\ map shows highly symmetric north and south 
spiral arms, whereas the northern spiral arm is not prominent at 
optical and FUV wavelengths.

Construction of various data products such as molecular gas and SFR
surface density maps and the associated error maps are summarized in
Appendix \ref{appen:sfrmap} and \ref{appen:gasmap}.  
The surface density maps are de-projected using the inclination angel 
$i\sim30^{o}$.
In this study we construct four different SFR surface density maps 
from combining FUV and 24 \um\ (\Sfuv), extinction corrected \halpha\ 
(\Shal), observed \halpha\ and 24 \um\ (\Scom), and 24 \um\ (\Smir) 
following various prescriptions in the literature.
All four of SFR surface density maps are expressed in units of 
\msunyrpc. The limiting $1\sigma$ sensitivity in surface density 
varies from map to map where the \SFRcom\ map has the highest 
rms sensitivity ($\sim0.1$ \msunyrpc) among all four SFR tracer 
maps. We adopt this value as the limiting sensitivity for all \Ssfr\ 
maps. 

The inclination corrected limiting surface densities corresponding 
to the sensitivity limits (Table \ref{image_table}) of the CARMA 
interferometer and the IRAM single dish maps are $\sim5.3$ \msunpc\ 
and $\sim0.6$ \msunpc, respectively. As mentioned in section 
\ref{stingdata}, we apply a multiplicative factor of 4 to the 
sensitivity limit of the IRAM single dish map to derive the limiting 
surface density. We combine the interferometric and single-dish maps 
to create the molecular gas surface density map (\Shtwo). This combined 
map finally convolved with a Gaussian beam to the obtain the spatial 
resolution of 6\arcsec. The typical $1\sigma$ sensitivity of this 
\Shtwo\ map is $\sim$3.7 \msunpc\ (inclination corrected). We study 
the molecular gas-SFR surface density relation for 
$\Shtwo \sim 10-245$~\msunpc.

\subsection{Sampling and fitting considerations}
Since one of the goals of this study is to explore how the functional
form of the SF law depends on the treatment of the data, we analyze
the images using 1) pixel analysis, incorporating all the data above 
a signal-to-noise cut, 2) aperture analysis, where we average over
circular apertures selecting bright regions, and 3) azimuthally
averaged annuli, with a width of 500 pc.

In local SF law studies, especially for normal star-forming galaxies
such as NGC~4254, the dynamic range probed by the molecular gas is
rather small. The observed dispersion in the SFR tracer, on the other
hand, is usually quite large depending on the selection of the SFR
tracer. Due to this characteristic of the gas-SFR surface density
relation, the determination of the functional form of this relation
depends critically on statistical methodologies and fitting
procedures. We describe the sampling and fitting strategies in detail
in the appendix sections \ref{appen:sample} and \ref{appen:stat}.

\section{Diffuse Extended Emission}
\label{back} 

Several components contribute to the total emission in galaxies.
Images contain emission from backgrounds or foregrounds, which are 
not physically related to the galaxy. Besides the emission of the
localized SF, they also contain diffuse components that are extended
over the entire disk and not necessarily associated with SF activity,
which we discuss in the following section. Since the calibration of
SFR tracers is frequently performed in star-forming regions, it may 
be important to remove the contribution from diffuse components to 
the brightness distribution before interpreting it in terms of a SFR.

The CO distribution of a galaxy can also contain diffuse emission not
necessarily associated with the individual star-forming regions. A
collection of unresolved small molecular clouds, in particular
Taurus-like clouds in the Milky Way with masses $\rm M \sim 10^4$ \msun, 
will fall below our detection threshold as individual entities but 
would contribute to the diffuse extended emission. It should be removed 
if those clouds do not host massive SF contributing to the SFR tracers.

\tabletypesize{\scriptsize}
\begin{deluxetable*}{cccccc}[h!]
\tablewidth{0pt}
\setlength{\tabcolsep}{0.15 in} 
\tablecolumns{6}
\tablecaption{Diffuse Fraction at Various Filter Scales}
\tablehead{
\colhead{Number}   
&\colhead{Filter Width}   
&\colhead{Filter Width} 
&\colhead{}  
&\colhead{Diffuse Fraction}  
&\colhead{}       \\
\cline{4-6}
\colhead{}
&\colhead{(arcsec)}
&\colhead{(kpc)}
&\colhead{\ffuv}  
&\colhead{\fdig}  
&\colhead{\fmir}       
} 
\startdata
I     & 75  &6.03   &0.72  &0.55  &0.68 \\
II    & 90  &7.23   &0.69  &0.50  &0.62 \\
III   &105  &8.44   &0.66  &0.45  &0.56 \\
IV    &120  &9.64   &0.62  &0.40  &0.50 \\
V     &135  &10.85  &0.59  &0.35  &0.44 \\
VI    &150  &12.06  &0.55  &0.30  &0.38 \\
VII   &165  &13.27  &0.52  &0.25  &0.34 \\
VIII  &180  &14.47  &0.48  &0.21  &0.30 \\
IX    &195  &15.68  &0.45  &0.18  &0.25 \\
X     &210  &16.88  &0.42  &0.15  &0.22 \\
XI    &225  &18.10  &0.39  &0.11  &0.19 
\enddata 
\tablecomments{Diffuse fractions as a function of 
galacto-centric radius that area obtained from 
different SFR maps are shown in Fig. \ref{de} in 
the appendix \ref{appen:mask}. }
\label{df_table}
\end{deluxetable*} 

\subsection{Diffuse Emission in the SFR Tracers} 
\label{DESFR}

The DE is ubiquitous in the UV, \halpha\, and 24
\um\ maps and it spreads out across the disk over a few tens to
hundreds of parsecs from the clustered OB association and resolved
\hii\ regions. The origin of this emission an active area of research 
over the past four decades (see Monnet 1971; Reynolds \etal\ 1971; 
Haffner \etal\ 2009 and references therein for diffuse emission from 
\hii\ regions; see Witt 1968; Hayakawa \& Yoshioka 1969; Meurer \etal\ 
1995; Pellerin \etal\ 2007 and references therein for diffuse UV 
emission; see Dale \etal\ 2007; Draine \etal\ 2007; Verley \etal\ 2007, 
2009 for diffuse 24 \um\ emission).

The extended \halpha\ emission originates in diffuse ($n_e\sim0.1$
cm$^{-3}$) warm (T$\sim$8000 K) ionized gas (DIG), which is analogous
to the warm ionized medium (WIM) in our own galaxy.  This gas has a
volume filling factor of $\sim$0.25 and scale height of $\sim$1 kpc 
and is a major component of the interstellar medium of the Galaxy (see 
Reynolds 1991, 1993 for reviews). In the Milky Way the DIG contributes 
$\sim10-15\%$ to the total \hii\ emission.
For external galaxies, however, observational evidences suggest that 
the DIG may contribute a substantial fraction ($\sim$30-60\%) to the 
total emission, fairly independent of galaxy Hubble type, inclination, 
and SF properties 
(Hunter \& Gallagher 1990, 1992; Rand \etal\ 1990;  Walterbos 
\& Braun 1994; Kennicutt \etal\ 1995; Hoopes \etal\ 1996, 2001; 
Ferguson \etal\ 1996; Hoopes \& Walterbos 2000; Wang \etal\ 1997; 
Greenawalt \etal\ 1998). Using spectral data Blanc \etal\ (2009) 
very recently reported 11\% DIG contribution in the central 4 kpc in 
M51. They also find that the DIG makes 100\% of the total emission 
coming from the interarm regions of this galaxy.

Is the emission from the DIG a tracer of SF ? On the one hand, the 
large energy output and the morphological association of the DIG 
with \hii\ regions have been used to argue that early type OB stars 
in \hii\ regions are the sources of this diffuse emission. Leakage 
of ionizing photons from porous \hii\ regions has been invoked
to explain the widespread distribution of this component.  This
requires the interstellar medium (ISM) to have low FUV extinction
along certain lines-of-sight, allowing a large mean free path for
these photons likely through interconnecting ionized bubbles 
(Tielens 2005, Seon 2009). On the other hand,
observational studies suggest that the DIG may not be entirely
associated with the early-type massive OB stellar clusters in
\hii\ regions. A population of late-type field OB stars (Patel \&
Wilson 1995a,b; Hoopes \etal\ 2001) or supernovae shocks may also
provide energy to the DIG (Collins \& Rand 2001; Rand \etal\ 2007).
The relative contribution of each of these sources to the DIG energy
balance is uncertain. Recent numerical simulations, however, suggest
that the contribution from \hii\ regions to the energy budget of the
DIG could be $\sim$30\% or less (Seon 2009).

Because mostly non-ionizing photons contribute to it, the diffuse UV
continuum emission has an origin different from that of the DIG. In
starburst galaxies, Meurer \etal\ (1995) found that about 80\% of the
UV flux at 2200 \AA\ is produced outside clustered OB associations 
and it has an extended character.  Popescu \etal\ (2005) suggested UV
light scattered by dust as the possible origin of the diffuse UV
emission. Tremonti \etal\  (2001) and Chander \etal\ (2003, 2005),
however, have noted that for starburst galaxies the spectral UV lines
from clusters are different from those in the inter-cluster
environment. Their studies show that the UV stellar signature in
clusters is dominated by O-type stars, while the inter-cluster
environment is dominated by less massive B-type stars. 
Late type OB field stars were also suggested by Hoopes \etal\ (2001)
as the origin of the diffuse UV emission in normal spirals.  In a
recent study, Pellerin \etal\ (2007) find that $\sim$75-90\% of the UV
flux is produced by B-type field stars in the disk of the barred
spiral NGC 1313.  These studies suggest that B-type field stars are
the major source of non-ionizing UV emission in galaxies, with a much
smaller contribution from scattered OB cluster light. This implies that 
the SF history has an important role in determining the ratio between 
the diffuse UV continuum and that arising in compact OB associations.  
Late B-type stars are longer lived ($\sim100$ Myr) and less massive 
($\sim5-20 $\msun) than O-type stars (age $\sim$10 Myr, mass 
$\geq 20$ \msun), with the latter types mostly residing in clustered 
associations.

The 24 \um\ continuum emission also has a diffuse component associated
with it.  In galaxy disks, 24 \um\ dust emission is frequently found 
near discrete \hii\ regions (Helou \etal\ 2004). This extended 24 \um\ 
emission is due to small dust grains out of equilibrium with the 
radiation field, for which single-photon events produce large 
temperature excursions (Desert \etal\ 1990). In addition to this 
localized emission, 24 \um\ sources are surrounded by a diffuse 
component associated with overall galaxy profile and internal structure 
such as spiral arms (Helou \etal\ 2004; Presscott \etal\ 2007; Verley et 
al. 2007, 2009). The old stellar population is thought to be responsible 
for such component, which comprises $\sim30-40\%$ of the total thermal 
dust emission in the central regions to $\sim60-80\%$ of the integrated 
emission in the extended disk (Draine \etal 2007; Verley \etal\ 2009).  

Understanding the nature and sources of DE is of great importance in 
studies of SF. Kuno \etal\ (1995) and Ferguson \etal\ (1996) discussed the 
role of the diffuse component when deriving the SFR based on \halpha\ 
emission. An assessment of the magnitude of DE contribution is necessary 
in order to use the SFR tracers derived from FUV, \halpha, and MIR 24 \um\ 
dust emission. Thus, it is interesting to study the contribution of the DE 
in these tracers and its consequences on spatially resolved molecular gas 
SF law studies. The contribution from DE is most important in the low 
surface brightness regime, where it can flatten the power law index of 
the SF law if unaccounted for.

\subsection{Diffuse Emission in the Molecular Gas}
\label{DECO}
In the Milky Way most CO emission is associated with GMCs, which have
a top heavy mass function (most of the mass is in the most massive
GMCs; Solomon \etal\ 1987). Similar ``top heavy'' GMC mass functions
are observed in most Local Group galaxies with the exception of M33
(Engargiola \etal\ 2003).
Even in the Milky Way, however, there exists a population of CO-emitting 
molecular clouds that are considerably more diffuse and have lower masses 
and column densities than GMCs that host massive SF. Examples 
are the high latitude clouds (Magnani \etal\ 1985), with typical column 
densities of $\rm N_{H2} \sim 10^{21} \ cm^{-2}$ and very low SF activity.

We do not know whether the GMC mass function in NGC~4254 is top heavy 
or bottom heavy. Even if it is top heavy, at 500 pc resolution our 
observations (1$\sigma \ \Shtwo \sim 3.7 \ \msunpc$) will be sensitive 
only to GMCs with masses $\rm M \gtrsim 10^6$
\msun\ as distinct entities. Localized, lower mass GMCs will be blended 
together and will appear as a blurred diffuse emission background in the 
galaxy disk. If these lower mass GMCs host no massive SF contributing to 
the selected SFR tracer they should not be included in the determination 
of the SF law, otherwise their inclusion will artificially steepen the 
power law index of the SF law.  

\subsection{Removing Diffuse Emission}
We will discuss first the treatment of the DE in the SFR tracers. 
The procedure for removing the DE in the molecular gas map is very 
similar, and is discussed in \S\ref{DEinCO}.

There is no standard procedure to separate integrated flux into emission 
from discrete star-forming regions and an extended diffuse component. 
Various criteria have been adopted in previous studies to distinguish 
between the discrete \hii\ regions and the DIG in the nebular emission 
map. For example, forbidden line ratio of \sii\ $\lambda \lambda$6716, 
6731 to \halpha\ (Walterbos \& Bruan 1994), the equivalent width of 
\halpha\ line (Veilleux \etal\ 1995), surface brightness cut (Ferguson 
\etal\ 1996), unsharp masking (Hoopes \etal\ 1996), \hii\ region 
luminosity function (Thilker \etal\ 2002).
Although the forbidden line ratios are powerful probes to separate the
components unambiguously, in absence of such information
two-dimensional (2D) spatial filtering techniques such as unsharp
masking is the most robust method to obtain information about DE.

We use unsharp masking to model and remove the DE, taking advantage of
its extended nature.  Several authors have used unsharp masking to
separate DE and discrete \hii\ regions in Local Group galaxies (Hoopes
\etal\ 1996; Greenawalt \etal\ 1998; Thilker \etal\ 2005). Our
approach, slightly different than these studies, is simple, easy to
implement, and avoids an ad hoc surface brightness cut used in other
studies.  The process involves creating a smoothed or blurred image
produced by a 2D moving boxcar kernel (middle panels,
Fig. \ref{ummap}) and then the subtraction of the smoothed map from
the original image (see Appendix \ref{appen:mask} for a more detailed
explanation). It is possible to use other kernels (Gaussian, Hanning,
etc), but boxcar is the computationally simplest.  The resulting final
image (lower panels, Fig. \ref{ummap}) has a reduced contribution from
the background as well as the DE, since most of it is contained in the
spatially smoothed map.  Indeed deep \halpha\ images of the
distribution of DIG in the local group members, such as M~33 and M~81,
show it to be quite smooth (Greenawalt \etal\ 1998).

Since we are using multi-wavelength SFR tracers, it is important 
to understand the nature and brightness of the DE as observational 
studies suggest that these properties depend on the wavelength studied. 
For example, studies of the fraction of DE at 24 \um\ in the disk of 
M~33 find $\fmir\sim 60-80\%$, increasing radially outward (Verley 
\etal\ 2009).  By contrast, the DIG shows the opposite trend, with 
$\fdig\sim 60\%$ at the center and decreasing to almost zero towards 
the outer disk. It is generally near 40\% across the disk (Thilker 
et al 2005). The diffuse fraction in FUV shows a remarkably flat 
profile $\ffuv\sim0.65$ (Thilker \etal\ 2005).

The crucial aspect of unsharp masking is the choice of the size of the
median filter kernel. The filtering kernel size affects the fraction
of the total emission of the original map contained in the smoothed
image, \fde, with the larger fractions in the smooth or diffuse
component corresponding to the smaller filtering kernel sizes.  In the
subsequent analysis we will use \fde\ to refer the diffuse fraction in
general. To refer to the diffuse fraction in the FUV, H$\alpha$, and
24 \um\ we use the notation \ffuv, \fdig, and \fmir, respectively. 

The details of the unsharp masking process can be found in the appendix 
\ref{appen:mask}. In NGC~4254 we explore a number of filter sizes in 
each SFR tracer, carrying out our analysis for each case (see Table 
\ref{df_table}). The diffuse fraction as a function of filter scale 
is shown in the panel (D) in Fig. \ref{de}. 
At a given filter scale the diffuse fraction is different for different 
SFR tracers. For convenience of presentation, therefore, we use \fdig\ 
as the reference DE since it is widely known in the astronomical 
community. We will refer the DE as the dominant, significant, sub-dominant, 
and negligible part of the total disk emission in the H$\alpha$ map for 
$\fdig \gtrsim 50\%$, $\sim$30-50\%, $\sim$10-30\%, and $\lesssim 10\%$, 
respectively. 

It should be noted here that the noise in the spatially smoothed map 
is negligible compared to the original image. The subtraction of the 
smoothed map, therefore does not change the noise properties of the 
original map.

\section{Results and Discussion}
\label{result}                  

We begin our analysis presenting azimuthally averaged radial
distributions of surface densities to demonstrate their spatial
variations. Fig. \ref{dist} shows the derived surface densities of the
SFR tracers as well as CO gas maps at $\fde=0$. Each panel shows the
1$\sigma$ dispersion in the radial distributions out to $0.6\optrad$,
where $\optrad\sim12.5$~kpc is the optical radius of NGC~4254.

The figure also shows the radial profiles of molecular gas obtained
from the interferometer-only observations and from the combined
single-dish/interferometer data. Both maps are in excellent agreement
within the central 80\arcsec\ ($\sim6.5$ kpc), showing that the 
interferometric data by itself accurately traces the distribution of 
cold molecular gas in the inner $\sim6.5$ kpc of this 
galaxy. The combined map, on the other hand, traces better the low 
surface brightness CO in the outer regions of the disk

\subsection{The SF Law using Pixels and Apertures}
\label{funcform}
In this section we present our results for the case when DE is
subtracted only from the SFR tracer maps. A more general case which
addresses DE in both the SFR and CO maps will be presented in the
following section.  The Nyquist sampling rate at the fixed $2.5\sigma$
cut results in $\sim 800$ approximately independent pixels for both
gas and SFR surface density maps at the dominant diffuse
fractions. The number of pixels increases to $\sim 950$ when the
contribution of DE is sub-dominant or negligible, because fewer pixels
fall below the signal-to-noise cut after DE subtraction.

We show our results for the \Ssfr\ - \Shtwo\ relation in
Fig. \ref{pixdist}. The figure shows the gas-SFR surface density
relation for pixel sampling at various \fde. The gray scale represent
the two dimensional histogram of the frequency of points, and the
contours are placed at 90\%, 75\%, 50\%, and 25\% of the maximum
frequency. The diagonal dotted lines represent lines of constant SF
efficiency ($\epsilon$), or constant molecular gas exhaustion
timescale (\tdep) with values of $\epsilon=1\%$, 10\%, and 100\%
corresponding to exhaustion times of $\tdep=10$, 1, and 0.1 Gyr
respectively. The filled circle in each panel represents the disk
averaged surface densities measured within \optrad\ before unsharp
masking. Within the range of diffuse fractions probed the \Smir\ -
\Shtwo\ relation shows the tightest correlation whereas the \Shal\ -
\Shtwo\ relation shows the largest scatter. We compute the linear
Pearson correlation coefficient ($\rm r$) for these two relations in
the range of explored diffuse fractions, finding $\rm r \sim0.55-0.7$
for the former and $\rm r \sim0.4-0.55$ for the latter. The observed
dispersion is $\sigsct \sim0.3$ dex for \Smir\ - \Shtwo\ and
$\sim0.5$ dex for \Shal\ - \Shtwo. 

Note that the scatter in the SF law is substantially lower when no DE 
is subtracted from the total emission of the SFR tracers. Furthermore, 
since the DE is proportionally more important in fainter regions, its 
subtraction increases scatter in the gas-SFR relation mostly at low 
surface densities. For the same reason, removing the DE steepens the 
SF law.

The results of aperture sampling are shown in Fig. \ref{aptdist} for 
a 105$^{\arcsec}$ unsharp masking kernel. The distributions of points
are overlaid on the contours obtained from the pixel analysis. By
construction the apertures sample mostly the high density regions, and
the overall agreement in these regions is excellent between the pixel
and the aperture analysis. The lack of the low surface brightness tail
along the vertical axes, however, has important implications for the
slope of the SF law, as we discuss next.

We show the measurements from different bivariate regression methods
in Fig. \ref{pixfit}. The figure highlights the \Scom\ - \Shtwo\
relation for pixel sampling at 105\arcsec\ filter scale. The scale
corresponds to a case of significant to dominant diffuse fraction,
\fdig=0.45\ and \fmir=0.56. The figure shows that the FITEXY method
yields the shallowest slope. For this gas-SFR surface density relation
the power-law index is in the range $\nmol \sim1.3-1.6$ for pixel
sampling, with intrinsic scatter $\sigsct
\sim0.3-0.4$ obtained from the fits. All fitting methods yields
systematically shallower slope ($\nmol \sim0.8-1.2$) and smaller
scatter ($\sigsct \sim0.1-0.3$ for both aperture sizes) for aperture
sampling.

Our exploration of the range of results obtained for the SF law by
using different methodologies is summarized in Fig. \ref{ind}. The
figure shows the dependence of the power-law index of the SF law on
the subtracted diffuse fractions. The vertical bars associated with
each point are purely methodological, and show the range of slopes
obtained from applying the different fitting methods described in
appendix \ref{appen:stat}.  We refer to them as the methodological
scatter, \sigfit. The mean (\nmolfit) obtained from averaging the
results of the three fits for each \fde\ is shown with filled circles.
The panels in this figure illustrate how the measurement depends not
only on the chosen SFR tracer, but also on the type of analysis and on
the treatment of the DE. The linearity of the functional form of the
molecular SF, in particular, depends on the amount of DE assigned to
either axis. It is important to notice, however, that this slope
change is driven by the lowest surface brightness regions of the
disk. In the high surface-brightness regions sampled in the aperture
analysis the choice of $\fde$ is unimportant, and a unique slope is
consistent with the data for any (reasonable) amount of DE. In these
regions the SF law is approximately linear, although its precise 
value depends on the SFR tracer.

We observe a direct relation between the slope of the SF law and the
magnitude of the DE subtracted in the pixel analysis (left panels,
Fig. \ref{ind}).  For a dominant diffuse fraction
($\fde\gtrsim50-60\%$ of the total disk emission), all the resolved SF
law relations in the pixel analysis show systematically the steepest
power-law indices, $\nmolfit \sim1.4 - 1.7$. For sub-dominant to
negligible diffuse fraction ($\fde\lesssim 30\%$ of the total disk
emission), however, the slope clearly becomes shallower, $\nmolfit
\sim1-1.2$. Thus higher \fde\ corresponds to a steeper power-law
index. This is only observed in the pixel analysis, which contains the
low surface brightness regions. Furthermore, the scatter in the
results yielded by the different fitting algorithms is also a
monotonically increasing function of the amount of DE subtracted. This
methodological scatter is driven by the corresponding increase in the
scatter of the low surface brightness pixels, which have a very broad
distribution for large \fde.

For the aperture analysis the fitted power indices are systematically
shallower than for pixel analysis, and robust to the choice of \fde\
and the aperture size (right panels in Fig. \ref{ind}). For a dominant 
diffuse fraction ($\fde\gtrsim 50\%$), the measurements of the SF law 
slope are consistent with $\nmolfit \sim1$. For small 
diffuse fractions ($\fde\lesssim 20\%$) diffuse fraction, the fitted 
power-index is in the range $\nmolfit \sim0.8 - 1$.  
Furthermore, the \halpha\ corrected for azimuthally averaged extinction 
tends to consistently have the steeper slopes (and the highest 
methodological scatter).  Although with this data sampling the slope 
still flattens monotonically with the reduction of the amount of 
subtracted DE, the dependence on \fde\ is very weak 
and the variation in \nmolfit\ is within the scatter 
of the different fitting methods. 

The power law index is slightly steeper for the larger, 1 kpc diameter
apertures at the dominant and significant diffuse fractions where the
fitting procedures diverge more from one another. This is likely due
to a combination of the fact that the larger apertures encompass some
area of low surface density material, and to the reduction in the
number of data points by a factor of $\sim3$. With fewer data the
regression analysis becomes highly sensitive to the distribution of
points. For sub-dominant to negligible diffuse fractions ($\fde
\lesssim 0.3$), however, the power law index of the local SF agrees
well for both aperture sizes.

Tables \ref{pixel_table} - \ref{apt2_table} show the derived parameters 
for median filter of size 75\arcsec, 105\arcsec\ and 180\arcsec.
The row represented by dash in each table show the parameter for \fde=0, 
i.e., when the filter size is the same of the entire map. The filter 
widths are chosen to show the representative cases of the dominant, 
significant, sub-dominant, and negligible diffuse fractions. At a given 
filter scale the estimates from the OLS bisector, FITEXY, and LINEXERR 
methods are shown by the top, middle, and, bottom row, respectively.
The quoted error in each parameter comes from bootstrap sampling of 
1000 realizations of data points.

\subsubsection{The SF Law in Annuli}
Many of the early resolved studies of the relation between gas and
star formation in galaxies analyzed the data using azimuthal averages
(e.g., Wong \& Blitz 2002).  Following the procedure similar to that
discussed in appendix \ref{appen:sample} to select the common regions 
from the $\Shtwo$ and $\Ssfr$ maps, we also explore the SF law for
azimuthally-averaged radial profiles (Fig. \ref{radial}). Sampled in
this manner, the functional form of the SF law in NGC~4254 is linear
($\nmolfit\sim1$) for $\fde=0$, and approximately linear ($\nmolfit
\sim1-1.2$) in the range of diffuse fractions studied. The linear form
stems from the fact that the azimuthal averages are dominated by the
high surface brightness regions, and there is no ``extended tail'' of
low $\Ssfr$ points steepening the fit to the distribution.  Since the
data have low dispersion all fitting methods yield consistent results. 
Table \ref{radpro_table} shows the fitted parameters derived from the 
OLS bisector method in all SFR tracers at the diffuse fractions 
highlighted in Fig. \ref{radial}.

\subsection{Dispersion in the Relations}
The intrinsic dispersion (\sigsct; see appendix \ref{appen:stat}) in 
the gas-SFR surface density relations at various diffuse fractions is
shown in Fig. \ref{scatter}. The figure shows the mean (\sigsctfit)
and the scatter (``dispersion of dispersion'') obtained by the three
regression methods. The SFR obtained from 24 \um\ displays the
tightest correlation with the molecular gas, among all tracers
($\sigsct \sim0.1-0.3$ dex). 
This is likely due to a combination of two effects: 1) young
star-forming regions that are still embedded in their parent clouds
will emit brightly at 24 $\mu$m. Bright H$\alpha$ emission will only
happen when the HII is older and the parent cloud is at least
partially cleared (see also Helou et al. 2004; Relano \& Kennicutt
2009). And, 2) by its nature, this SFR tracer does not need to be
corrected by extinction. The spatial correspondence between the 24
$\mu$m and CO maps is striking (Fig. 2).

The extinction-corrected \Shal, on the other hand, shows the 
largest scatter ($\sigsct \sim 0.3-0.6$ dex) of all tracers, 
with or without unsharp masking. This is because the 
extinction correction is azimuthally averaged, and it does a poor 
job at correcting any position although it yields the correct 
result in a statistical sense. Using one galaxy-wide correction 
factor to remove the \nii\ $\lambda \lambda$6548, 6583 forbidden 
line emission from the \halpha\ map is also another potential 
(likely minor) contributor to the scatter, since it may well vary 
with the position.

Due to its large dispersion, the results for \Shal\ from the 
different regression methods differ substantially from one another 
($\sigfit \sim0.4$). By contrast the combined \SFRcom\ tracer, 
which applies the same underlying extinction correction locally, 
yields a tighter correlation ($\sigsctfit \sim0.1-0.5$ dex) and a 
flatter slope ($\nmolfit \sim 1-1.6$). The composite \SFRfuv\ 
yields very similar results to \SFRcom. The observed scatter also 
becomes somewhat smaller for larger apertures (dashed-dot line in 
the right panels of Fig. \ref{scatter}), particularly for \halpha\ 
which clearly benefits from averaging over larger regions.

\subsection{Diffuse CO Emission}
\label{DEinCO}
So far we have only considered the effect of diffuse emission, 
possibly unrelated to recent massive SF (on the vertical axis of the 
\Ssfr - \Shtwo\ plots). Should we be also concerned about analogous 
effects in the horizontal axis (\S\ref{DECO})? To explore the effect 
of removing a diffuse extended component from the CO distribution we 
use smoothing kernels of the same size in both the horizontal and 
vertical axes in molecular gas-SFR surface density plots.

Figure \ref{diffco} highlights the results for \SFRfuv\ and \SFRcom\ 
using pairs of lines to illustrate the methodological dispersion. 
The thin solid and dashed lines show the dependence of \nmol\ on \fde\
when both axes are subject to unsharp masking, at two different values
of the signal-to-noise threshold for including points. To serve as
comparison, the thick solid lines represents the case when only the
SFR tracer maps have undergone unsharp masking.  In most of our
analysis we have only included points where the gas surface density
map is $\Shtwo\geq2.5\sigma\sim9.2 \msunpc$ (dashed lines in
Fig. \ref{diffco}). To explore the effects of this threshold on the
analysis we also plot the results for $\Shtwo\geq1\sigma \sim3.7
\msunpc$ (thin black lines). The threshold for the SFR maps 
is kept at $2.5\sigma$.

This figure shows that our attempt at removing a diffuse molecular
component in NGC~4254 has only very mild impact on the results of the
analysis, and only for the pixel analysis. The results derived from
the apertures in the high surface brightness regions are essentially
unchanged. Interestingly, the consistency between the different
fitting methods is better than in the case where only the DE in SFR 
is removed. This is likely because errors in \fde\ subtraction smear 
the data along the main relation, rather than only in the vertical 
direction in gas-SFR surface density relation.

Lowering the threshold after unsharp masking the CO produces somewhat
flatter slopes at higher \fde, while increasing the dispersion of the
results. For example, the slope is approximately unity below
$\fmir\sim0.4$ for a \Shtwo\ cutoff value of 7.4 \msunpc, while for a
threshold of 10 \msunpc\ it would be unity only below $\fmir\sim0.25$.
The power-law index remains unchanged at the extremes of \fde.
The results for other two tracers are qualitatively similar to those 
presented in Fig. \ref{diffco}.
 
The dispersion in the gas-SFR surface density relation systematically 
goes down when both variables are subject to unsharp masking. We find 
$\sim10-20\%$ reduction in the scatter depending on the SFR tracer. 
The fitting methods tend to converge with one another because of the 
reduction in the scatter in the range $\fde \lesssim 0.3$, which is 
clearly evident in Fig. \ref{diffco}.

\tabletypesize{\small}
\begin{deluxetable*}{@{}l@{}@{}c@{}@{}c@{}@{}c@{}@{}c@{}@{}cc@{}}[h!]
\tablewidth{0pt}
\setlength{\tabcolsep}{0.1 in} 
\tablecolumns{6}
\tablecaption{Disk Averaged Parameters}
\tablehead
{\colhead{Param.}  
&\colhead{DE} 
&\colhead{FUV + 24 \um} 
&\colhead{\halpha}  
&\colhead{\halpha\ + 24 \um}  
&\colhead{24 \um}       
&\colhead{Unit} 
}
\startdata
SFR            &\fde=0     &$4.9\pm0.1$    &$2.4\pm0.1$   
&$3.5\pm0.1$   &$4.1\pm0.2$  &\msunyr \\
\Ssfr\         &\fde=0     &$13.3\pm1.2$   &$13.7\pm1.2$  
&$11.1\pm1$    &$10.5\pm1$ &\msunyrpc \\ \\ \cline{1-7} \\ 
$\tdep^{1}$    &max. \fde  
&$4.3\pm1.2$   &$5.5\pm1.9$   &$5.4\pm1.8$    &$4.7\pm1.5$  &Gyr \\
$\tdep^{2}$    &max. \fde  
&$2.0\pm0.5$   &$2.5\pm0.8$   &$2.4\pm0.7$    &$2.2\pm0.6$  &Gyr \\
$\tdep$        &\fde=0     
&$1.7\pm0.3$   &$3.3\pm0.8$   &$2.2\pm0.4$    &$2.0\pm0.3$  &Gyr
\enddata 
\tablecomments{The parameters are estimated within \optrad\ from the 
maps with zero subtraction of DE, i.e., maps with \fde=0. 
The $\tdep^{1}$ and $\tdep^{2}$ are estimates at maximum \fde\ when 
1) only the SFR tracer map, and 2) the SFR tracer and gas maps both 
are subject to unsharp masking.  The total molecular gas mass within 
\optrad\ is, $\rm M_{H_2, tot} = (5.1\pm0.9) \times 10^9$ \msun. 
The disk average gas surface density is, \Shtwo$=(46\pm3) \ \msunpc$.}
\label{dap_table}
\end{deluxetable*}

\subsection{Goodness of Fit}
How well does the power law functional form represent the SFR and
molecular gas surface density relationship analyzed in this study? A
measure of the goodness-of-fit of a model is to derive the $\chi^2$
statistic based on the least square method (Deming 1943). The best fit
lines provided by the FITEXY estimator have reduced-$\chi^2 \sim 1$
with probability $\sim0.35-0.5$. This is achieved by iteratively
adjusting the error along the Y-axis, $\rm \sigma_{y}^2 = \sigma_{m}^2 +
\sigma_{i}^2$, where $\sigma_{i}$ is the intrinsic scatter in the 
gas-SFR surface density relation and $\rm \sigma_{m}$ is the measurement 
error.

A graphical alternative to evaluate the goodness-of-fit is to test the
normality of residuals. The residuals are the deviations of
observational data from the best fit line. We perform this test for
the fitted lines produced by all three estimators. At large \fde\ the
distributions of residuals for various SFR tracers are approximately
consistent with a normal distribution, and they become more so when
\fde\ decreases.  Our analyses suggest that the observed relation
between the molecular gas and SFR surface densities in NGC~4254 is
consistent, at least to first order, with the power law form.

\subsection{Star Formation Efficiency} 
The star formation efficiency (\sfe) is a convenient, physically 
motivated way to parametrize the relationship between molecular gas 
and SFR. The SFE has been defined in various ways in literature. For 
example, it is defined as the ratio of the produced stellar mass to 
the total gas mass. This definition is more commonly seen in the 
Galactic studies (Myers \etal\ 1986) but also used in galaxy modeling 
(Vazquez-Semadeni \etal\ 2007).
For extra-galactic studies, the molecular gas \sfe\ is usually defined 
as (Young \& Scoville 1991; McKee \& Ostriker 2007),
\begin{equation}
\rm \sfe = \epsilon = \frac{\Ssfr}{\Shtwo} = 
\rm A \ \left[ \Shtwo \right]^{N_{mol}-1}.  
\label{sfe}
\end{equation}
\noindent The inverse of the SFE is considered as the gas depletion 
timescale, $\tdep=\sfe^{-1}$. This parameter is used to discern 
between the starburst and normal star-forming galaxies (Rowand \& 
Young 1999). For starburst galaxies the typical depletion time is 
hundreds of Myr whereas normal star-forming galaxies have depletion 
timescale of $\sim2$ Gyr (Bigiel \etal\ 2008; Leroy \etal\ 2008). 
We adopt the definition in Eq. \ref{sfe} for consistency with the 
studies of Bigiel \etal\ (2008), Leroy \etal\ (2008), and Blanc 
\etal\ (2009).

A linear functional relationship between the SFR and molecular gas
surface densities implies that the efficiency to turn molecular gas
into stars is constant across the disk. 
It also implies the gas consumption time is similar for both massive 
and low mass clouds. A linear molecular gas SF law
is consistent with the scenario in which GMCs turn their masses into
stars at an approximately constant rate, irrespective of their
environmental parameters. Observations suggest that GMCs properties
are fairly uniform across galaxies (Blitz \etal\ 2007; Bolatto \etal\
2008). 
A non-linear molecular gas SF law, however, implies that gas is 
turned into stars at a faster rate at higher surface densities. 

Figure \ref{sfede} shows the disk averaged molecular SFE (\tdep) as a
function of diffuse fraction. We derive the SFE finding the ratio of
SFR to molecular gas for each pixel or aperture in the map, and plot
the average with error bars computed from the standard deviation. The
SFE is robustly determined and independent of \fde\ when
$\fde\lesssim0.3$ for all SFR tracers. This is within the range of
ionized gas \fde\ observed in the Milky Way and Local Group galaxies
(e.g., Thilker et al. 2002), and also in recent spectroscopic
determinations in the central region of M~51 (Blanc et al. 2009). At
higher \fde\ the SFE changes (becomes lower) by up to a factor of
$1.5-3$ at the extreme. The global efficiency is approximately
independent of \fde\ (within 40\%) when both the SFR and the gas map
have a diffuse component removed (gray points), reflecting the fact
that the SF law is linear in that case.  At a given \fde, the global
SFEs derived from our four tracers are approximately consistent with
one another, although the SFE obtained from \halpha\ is only
marginally so (see Table \ref{dap_table}). The SFE in NGC~4254 is
essentially independent of radius up to $R\sim0.5\optrad\approx6$~kpc.
Inside $0.1\optrad$ the SFR derived from the extinction-corrected
\halpha\ image does not agree with the other SFR tracers, likely
because of a problem with the extinction correction.

Figure \ref{sfede} also shows that from sub-dominant to negligible 
\fde\ the molecular SFE in NGC~4254 is fairly typical of large spirals 
(see Table \ref{dap_table}). The disk averaged $\tdep\approx2.2\pm0.4$ 
Gyr estimated from \Scom\ map at $\fde=0$ is in good agreement with 
the $\tdep\approx2.1$ Gyr measurement by Wilson \etal\ (2009) in 
\cone, who used a similar SFR tracer.

\subsection{Systematics Affecting the Local SF Law}
\label{systematic}

\subsubsection{Effect of the Non-detections}
In this study we analyze regions that have values over the adopted
thresholds in both the \Ssfr\ and \Shtwo\ maps.  To check for the
effect of not including pixels that are detected in one axis but not
the other, we include every pixel out to \optrad. This results in
about 20 additional points, all with measurable SFR but no CO
detection and thus having only upper limits for their gas surface
densities. We find that these points closely follow the original
distribution of points detected in both \Ssfr\ and \Shtwo\ at the
limiting end of gas surface density. Thus, there are no new data
trends hidden in the limits. They comprise only 3\% of the total
number of points obtained with the data selection criteria as 
mentioned in appendix \ref{appen:sample}. The impact of these points 
on the determination of the functional form of the SF law is 
negligible.

\subsubsection{Variations in the Data Selection Cuts}
It is a common practice to adopt one specific sensitivity limit in 
analyzing the gas-SFR surface density relation (for example, 3$\sigma$ 
in Kennicutt \etal\ 2007; 2.5$\sigma$ in Bigiel \etal\ 2008; 2$\sigma$ 
in Verley \etal\ 2010). The reason for adopting these sensitivity cuts 
is to ensure the reliability of the data. The choice of sensitivity 
limit, however, may have an impact on the determination of the local 
SF law, particularly given the limited dynamic range of the data. To 
explore the effect of this choice we have analyzed the gas-SFR surface 
density relation at several thresholds above 2$\sigma$ sensitivity. We 
find that the choice of limit has a measurable effect on the slope of 
the SF law for the pixel analysis, such that lower thresholds steepen 
the slope by as much as $30\%-40\%$ (depending on the SFR tracer
considered), with a simultaneous increase in the dispersion of the low
surface brightness points. The determination of the slope for the
aperture sampling and the azimuthally averaged radial profile are, on
the other hand, robust to the choice of the sensitivity limit.

\subsubsection{Sensitivity to the Error Maps}
The results presented in this section are computed for a set of
measurement error maps of SFR and gas surface densities. These maps
are constructed under certain assumptions of the observational 
uncertainties. However, measurement uncertainties in flux calibration, 
continuum subtraction, and other parameters are propagated into the 
error maps. Variations in the assumptions made to include their 
contributions lead to changes in the error maps, which directly 
influence the regression analysis. For several sets of error maps with 
varying assumptions about the measurement uncertainties we find up to 
40\% variations in the slope measurements provided by the bivariate 
regression methods.

\section{Comparison With Previous Studies}
\label{compare}
In this section, we compare our results with recent studies of the 
spatially resolved SF law in nearby galaxies by Kennicutt \etal\
(2007), Bigiel \etal\ (2008), Blanc \etal\ (2009), and Verley
\etal\ (2010). While making comparison it should be borne in mind 
that, for a given SFR tracer and at a given kernel size, the table 
provides fitting results from three different bivariate regression 
methods. Our main results presented in various panels in Fig. 9, on 
the other hand, show the mean (\nmolfit) and the dispersion (\sigfit) 
of these three measurements.

Kennicutt \etal\ (2007) obtain a super-linear power law ($\nmol
\sim1.37\pm0.03$) and an observed scatter of $\sigsct \sim0.4$ dex 
for NGC~5194 (M~51) using \SFRcom\ as the SFR tracer, and apertures 
520 pc in diameter centered on arm and inter-arm star-forming regions 
of M~51. They find a somewhat shallower power-law index in larger
apertures (1850 pc in diameter). The authors subtract the diffuse
component contribution using measurements in rectangular regions of
the image, and employ FITEXY for the fitting. Although our aperture
placement method is not strictly identical since we place apertures on
the high surface density regions of NGC~4254, we find an approximately 
linear power law ($\nmolfit \pm \sigfit\sim0.9\pm0.1$) for our 
apertures independently of the DE subtraction in the SFR or molecular 
gas axis (see panel f in Fig. \ref{ind}). 
We also find that the functional form of the SF law is insensitive 
to the change of the aperture size, at least for the selected 
apertures in NGC~4254. The power law indices found at 500 pc and 1 
kpc diameter apertures are well within their respective error bars 
(see panel f of Fig. \ref{ind}). 
The observed scatter is systematically smaller ($\sigsctfit \sim0.1-0.3$ 
dex) in our study (see panel f of Fig. \ref{scatter}).

Bigiel \etal\ (2008) use the \SFRfuv\ tracer, similar methodology in
data analysis, OLS bisector fitting, and the same conversion factor,
\cf.  The authors find an approximately linear form 
($\nmol \sim0.96\pm0.07$) for the resolved molecular gas SF law
in a sample of star-forming disk galaxies at 700~pc resolution.
They measure a typical molecular gas depletion 
timescale  $\sim2$ Gyr, under the assumption that the DIG contribution 
is negligible in both the FUV and MIR maps. They find a typical scatter 
of $\sim0.2$ dex in the gas-SFR density relation. 
The authors did not consider the role of diffuse fraction in the 
SF law, therefore, their analysis is similar to the case $\fde=0$ 
studied in this paper. Comparing our results of pixel analysis at 
$\fde=0$, we find a very similar power law index 
($\nmolfit \pm \sigfit \sim 0.91 \pm 0.03$) in NGC~4254 (see panel 
a of Fig. \ref{ind}). The molecular gas depletion timescale for this 
galaxy is $\tdep\sim1.7$ Gyr. The estimate of intrinsic scatter 
obtained from OLS bisector method ($\sigsct \sim 0.2$ dex) is 
remarkably consistent with the Bigiel \etal\ (2008) study.  

Blanc \etal\ (2009) studied the central $\sim4$ kpc of M~51 using
spectroscopic observations to obtain $\Shal$.  They find a slightly
sub-linear ($\nmol \sim 0.82 \pm 0.05$) functional form of the SF 
law with an intrinsic scatter $\sigsct \sim 0.43$ dex. They place
apertures 170 pc in diameter and measure the DIG contribution using 
the \sii/\halpha\ ratio, finding $\fde\approx11\%$. Despite 
the different DE removal and sampling strategies, we also find
an approximately linear slope $(\nmolfit \pm \sigfit \sim 0.95 \pm
0.2)$ at a similar DIG fraction in NGC~4254 (see panel d of 
Fig. \ref{ind}).  The authors use a Monte
Carlo method for fitting the SF law, which treats the intrinsic
scatter in the relation as a free parameter. This method allows the
inclusion of non-detections in both the \Shtwo\ and \Ssfr\ maps, 
fitting the data in linear space. The advantage of this approach is 
that it is free from the systematics involved in performing linear 
regressions over incomplete data sets in logarithmic space. A drawback 
of this method is that it does not treat the data symmetrically, as 
it employs \Shtwo\ as the independent variable in the fits.  Bivariate 
statistical methods such as the ones in our study do not easily allow 
for the inclusion of upper limits. Nonetheless, they permit a robust 
parametrization of the data with a symmetric treatment of both axes, 
assuming that there is a good understanding of the data uncertainties.

Verley \etal\ (2010) study the local SF law in M~33 using
azimuthally-averaged radial profiles at a resolution of 240 pc, and
using apertures sampling at various spatial scales ($180-1440$~pc). 
These authors also did not consider the role of diffuse fraction in 
the SF law. They use two different fitting techniques, including 
FITEXY, and several SFR tracers, including extinction-corrected
\halpha, extinction-corrected FUV, and a combination of observed FUV
and total infrared luminosities. From aperture analysis they find
that the molecular gas SF law is always super-linear and steeper 
($\nmol \sim 1.4-1.8$) in \halpha\ than in the other SFR tracers 
($\nmol \sim 1.2-1.4$). They find the same trend at all spatial 
scales. Their radial profile analysis, on the other hand, shows a 
shallower power-law index ($\nmol \sim 1.1-1.3$) by all three 
tracers.  The results of our study are qualitatively similar to 
those of Verley \etal\ (2010). For example, our study also suggests 
that the power-law index is 
systematically steeper in \halpha. We also obtain a shallower SF 
law when using azimuthally-averaged radial profiles.  At coarser 
spatial resolutions, Verley \etal\ find a slight steepening of the 
SF law in all SFR tracers. However, we find little change in the 
power-law index over the explored range of spatial scales.

\section{Summary and Conclusions}
\label{summary}

We study the spatially resolved molecular gas SF law in NGC~4254
within the central $\sim6.5$~kpc, in an attempt to understand the
combined effects of the underlying assumptions on the functional form
of the relation between molecular gas and SF. To this end, we use four
different SFR tracers. These are \halpha\ with azimuthally-averaged
extinction correction, 24 \um, combined \halpha\ and 24 \um, and
combined FUV and 24 \um. We utilize various fitting procedures (the OLS 
bisector, FITEXY, and LINEXERR; described in appendix \ref{appen:stat}) 
to constrain the parameters of the local SF law. We explore the
effects of error weighting and signal-to-noise cuts on the results.
We employ three different sampling strategies (pixel-by-pixel, aperture, 
and azimuthal averages) to probe the gas-SFR surface density relation 
at various spatial resolutions (500 pc and 1 kpc) and surface densities.

We investigate the effect of diffuse emission on our ability to
measure the local SF law. Diffuse emission is an ubiquitous component
of the maps in FUV, \halpha, and 24 \um, which may not be associated
with star formation and comprises an unknown but perhaps significant
fraction of the total disk emission (in the Milky Way, the diffuse
emission in \halpha\ related to the DIG constitutes $10-15\%$ of the
total emission; Reynolds 1991).  Similarly, there may be an analogous
component of DE in the molecular gas axis of the SF law, comprised of
unresolved clouds that are too small to host massive star formation.
The contribution from DE is most important in the low surface
brightness regime. To extract the DE from the maps we use spatial
filtering (unsharp masking). 

We study the gas-SFR surface density relation for the molecular 
gas surface density range $\sim10-245$~\msunpc. This range is 
typical of normal star-forming galaxies but significantly smaller 
than starburst galaxies. The lower limit of molecular gas surface 
density is consistent with recent studies which suggest that the 
atomic to molecular phase-transition occurs in the ISM at surface 
densities $\rm \Sigma_{gas}\sim10$~\msunpc\ (Wong \& Blitz 2002; 
Kennicutt \etal\ 2007; Leroy \etal\ 2008)

Without additional data (for example, optical spectroscopy) we 
cannot establish the fraction of DE corresponding to the different 
SFR tracers in NGC~4254. Therefore, we take \fde\ as an independent
parameter and we explore the molecular gas-SFR surface density
relation for varying diffuse fraction. We find that in the high
surface brightness regions sampled by our aperture analysis (and
dominating our azimuthal averages) the value of \fde\ has little or 
no impact on the power-law index of the SF law, which is approximately
linear in NGC~4254 for all the SFR tracers considered ($\nmolfit\sim
0.8-1.1$; Fig. \ref{ind} right panels). Thus, the value of the
logarithmic slope of the SF law can be robustly established in the
high surface brightness regions of this galaxy.

When lower surface brightness regions are included in the analysis,
the removal of the DE, the choice of SFR tracer, and the choice of
regression analysis all have an impact on the determination of the 
SF law (Fig. \ref{ind} left panels).  For sub-dominant DE fractions 
in the SFR maps ($\fde\lesssim0.3$), almost all tracers agree on an
approximately linear SF law ($\nmolfit\sim0.8-1.1$). Removal of a DE
component in both the SFR and the CO maps pushes this linear regime 
to slightly larger DE fractions ($\fde\lesssim0.4$;
Fig. \ref{diffco}). When the DE is assumed to dominate the emission
($\fde\gtrsim0.5$) all tracers yield a super-linear SF law
($\nmolfit\gtrsim1.4$). The discrepant tracer at low $\fde$ is
\halpha\ with a radially-dependent extinction correction. Since this
is the SFR tracer that yields that largest scatter ($\sigsctfit
\sim0.4-0.6$ dex) in the gas-SFR surface density relation, we think
this steepness is an artifact of the combined effects of noise and
performing the regression in logarithmic space.

We explore a small range in spatial scales, as set by our choice of
aperture sizes (500 pc and 1 kpc). In this range, we find that the
results we obtain are independent of the spatial scale, with a very
slight tendency to recover a steeper (but still approximately linear)
SF law with a smaller scatter for larger aperture sizes
(Fig. \ref{ind} right panels, points and dash-dotted lines). On the
scales of the azimuthally-averaged radial profiles we recover an
approximately linear SF law for all \fde\ ($\nmolfit\sim1-1.2$;
Fig. \ref{radial}).

In NGC~4254 the intrinsic scatter (\sigsct) of the SF law varies with
the choice of the SFR tracer (Fig. \ref{scatter}). In particular, the
24 \um\ emission shows the tightest correlation with the molecular gas
surface density ($\sigsctfit \sim0.1-0.4$ dex). This is likely because
24 \um\ emission is closely correlated with embedded SF, still
associated with the parent molecular material. This suggests that 24
\um\ is a very good tracer of SFR on timescales similar to the
lifetimes of GMCs over spatial scales of several hundred parsecs. The 
combined \SFRcom\ tracer, which essentially applies a local extinction 
correction to \halpha, yields the second tightest correlation 
($\sigsctfit \sim 0.1-0.5$ dex).

The molecular SFE in this galaxy is constant out to $R\sim0.5\optrad$
(Fig. \ref{sfede}). The disk averaged depletion timescale of the
molecular gas ($\sim2$ Gyr) observed in NGC~4254 is fairly typical 
of large spirals (Bigiel \etal\ 2008; Leroy \etal\ 2008) and in good
agreement with previous observations (Wilson \etal\ 2009). The
different SFR tracers agree very well with each other except for
\halpha, which yields a somewhat longer molecular gas depletion time.
Like the power-law index, the SFE is independent of \fde\ when the DE
is sub-dominant. Removing a diffuse component from both the SFR and
the molecular gas yields a SFE that is independent of \fde.

Although the presence of diffuse emission not associated with star
formation in the tracer used to determine the SFR or the molecular
surface density should be a concern, at least in NGC~4254 the SF law
can be determined in a precise and robust manner with no exact
knowledge of the diffuse fraction in two cases: 1) in the high surface
brightness regions independent of the value of \fde, and 2) throughout
the disk if $\fde<30\%$. In both cases the resolved SF law is
approximately linear in this galaxy.

\acknowledgments
We thank the anonymous referee for many constructive comments. We thank 
Daniela Calzetti for useful suggestions and stimulating discussions. 
NR thanks Eric Feigelson and Brandon Kelly for suggestions on 
statistical methodologies; Danny Dale, Chad Engelbracht, Moire 
Prescott for SINGS data related issues; Johan Knapen and Rebecca 
Koopman for useful communications on \halpha\ images; Richard Rand for 
valuable suggestions regarding extended DE.
The authors thank the teams of SINGS and GALEX NGS for making their 
outstanding data set available. This research has made use of the NASA/IPAC 
Extragalactic Database (NED) which is operated by the Jet Propulsion 
Laboratory, California Institute of Technology, under contract with the 
National Aeronautics and Space Administration. We acknowledge the usage 
of the Hyper-Leda database (http://leda.univ-lyon1.fr). We have made use 
of NASA's Astrophysics Data System NASA/ADS.
Support for CARMA construction was derived from the Gordon and Betty 
Moore Foundation, the Eileen and Kenneth Norris Foundation, the Caltech 
Associates, the states of California, Illinois, and Maryland, and the 
National Science Foundation. Funding for ongoing CARMA development and 
operations are supported by the National Science Foundation (NSF) and 
the CARMA partner universities. This research is supported in part 
by grant NSF-AST0838178.

{\it Facilities:} \facility{{\em GALEX}}, \facility{KPNO}, 
\facility{{\em Spitzer}}, \facility{CARMA}, \facility{IRAM}.

\begin{figure}[t]
\epsscale{0.90}
\plotone{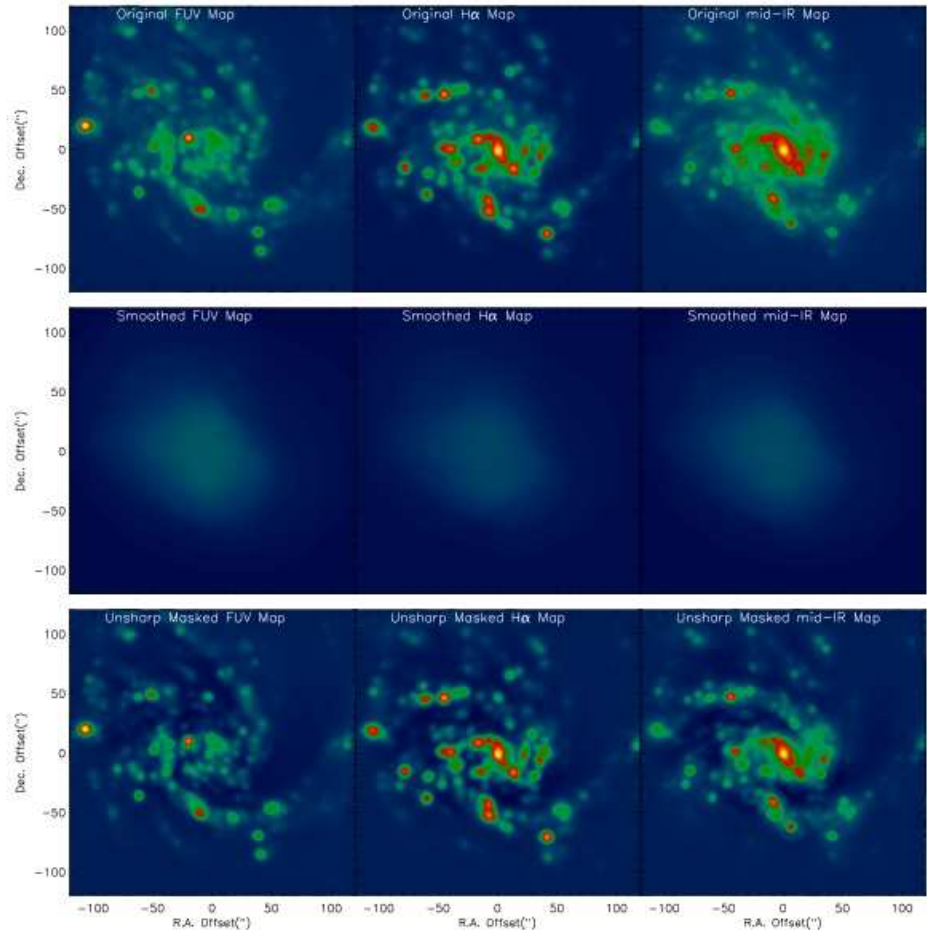}
\caption{Two-dimensional unsharp masking. 
The top panels show the original FUV, \halpha\ and MIR 24 \um\ emission 
maps from left to right. The middle panels show the corresponding smoothed 
images where the smoothing is performed by a 2D box kernel with the kernel 
size of 105\arcsec. The smoothed images are subtracted from the corresponding 
un-smoothed original images to produce unsharp masked maps (bottom panels). 
\label{ummap}}
\end{figure}

\begin{figure}
\epsscale{1.10}
\plotone{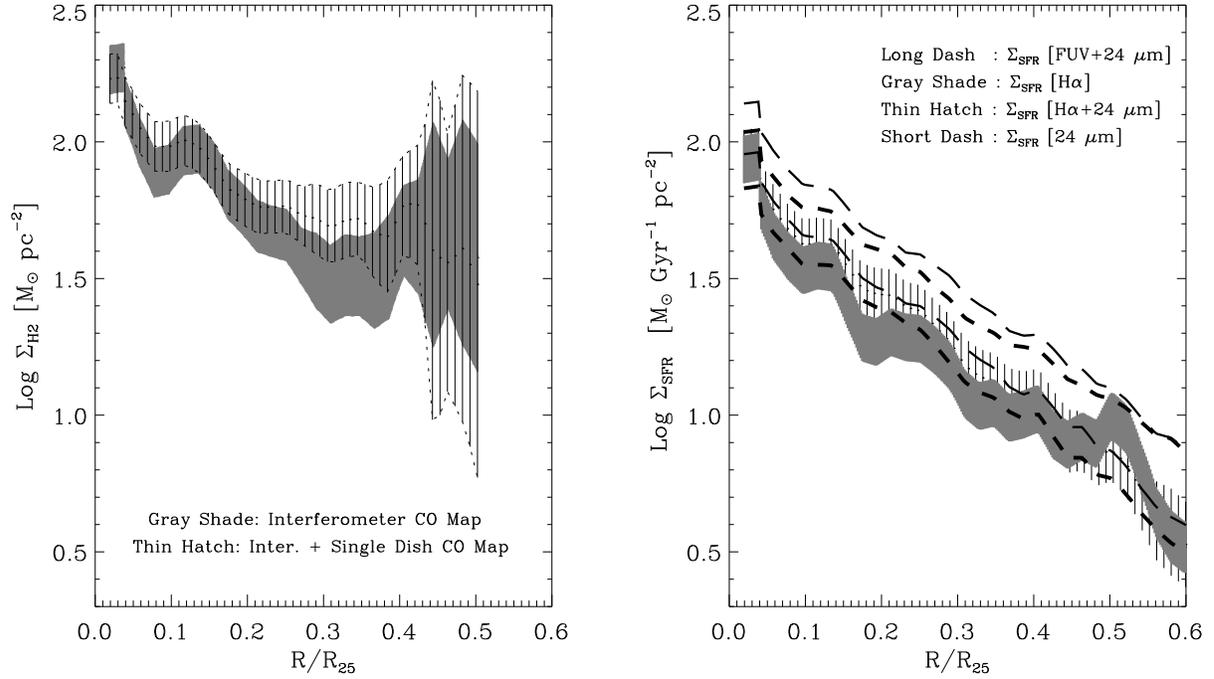}
\caption{Azimuthally averaged radial profiles of surface density maps 
shown at $1\sigma$ up and below the mean trend at each radial point. 
Left Panel: The radial distribution of molecular gas surface density is 
shown for the most sensitive CARMA interferometer \cone\ observation. 
The profiles derived from interferometer data alone as well as combining 
interferometer and single dish data are shown by gray shade and thin hatch, 
respectively. 
Right Panel: The radial profiles derived from \Sfuv, \Shal, \Scom, and 
\Smir\ maps are shown by long dashed line, thick hatched region, thin 
hatched region, and short dashed line, respectively. 
\label{dist}}
\end{figure}

\begin{figure}[t]
\epsscale{0.70}
\plotone{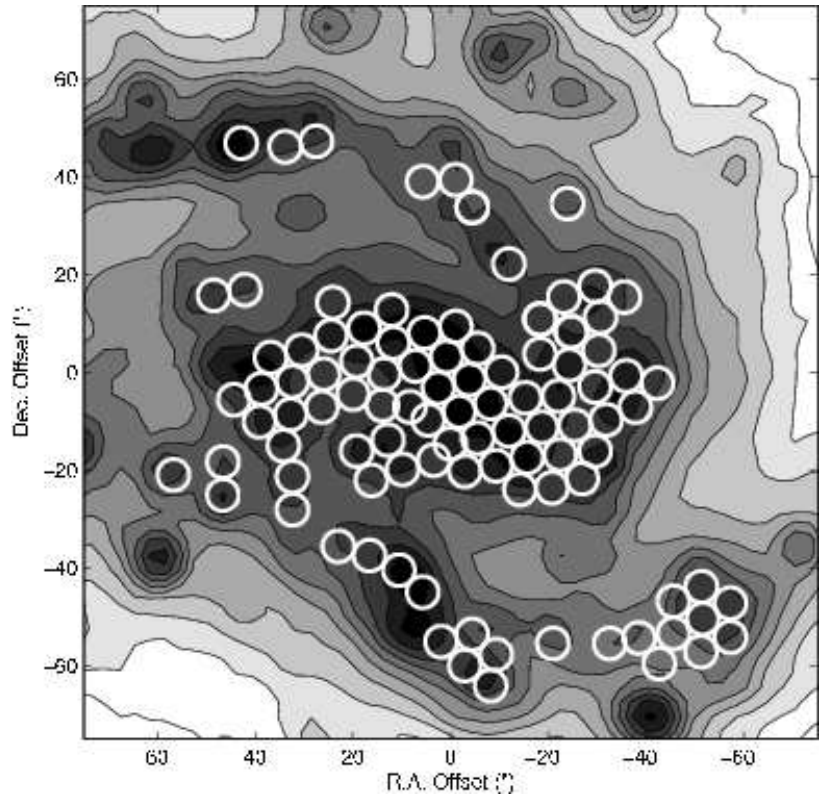}
\caption{The \Scom\ map with the 6\arcsec\ (500 pc) diameter apertures 
overlaid. The total number of such apertures is 102. 
\label{apt}}
\end{figure}

\begin{figure}[t]
\plotone{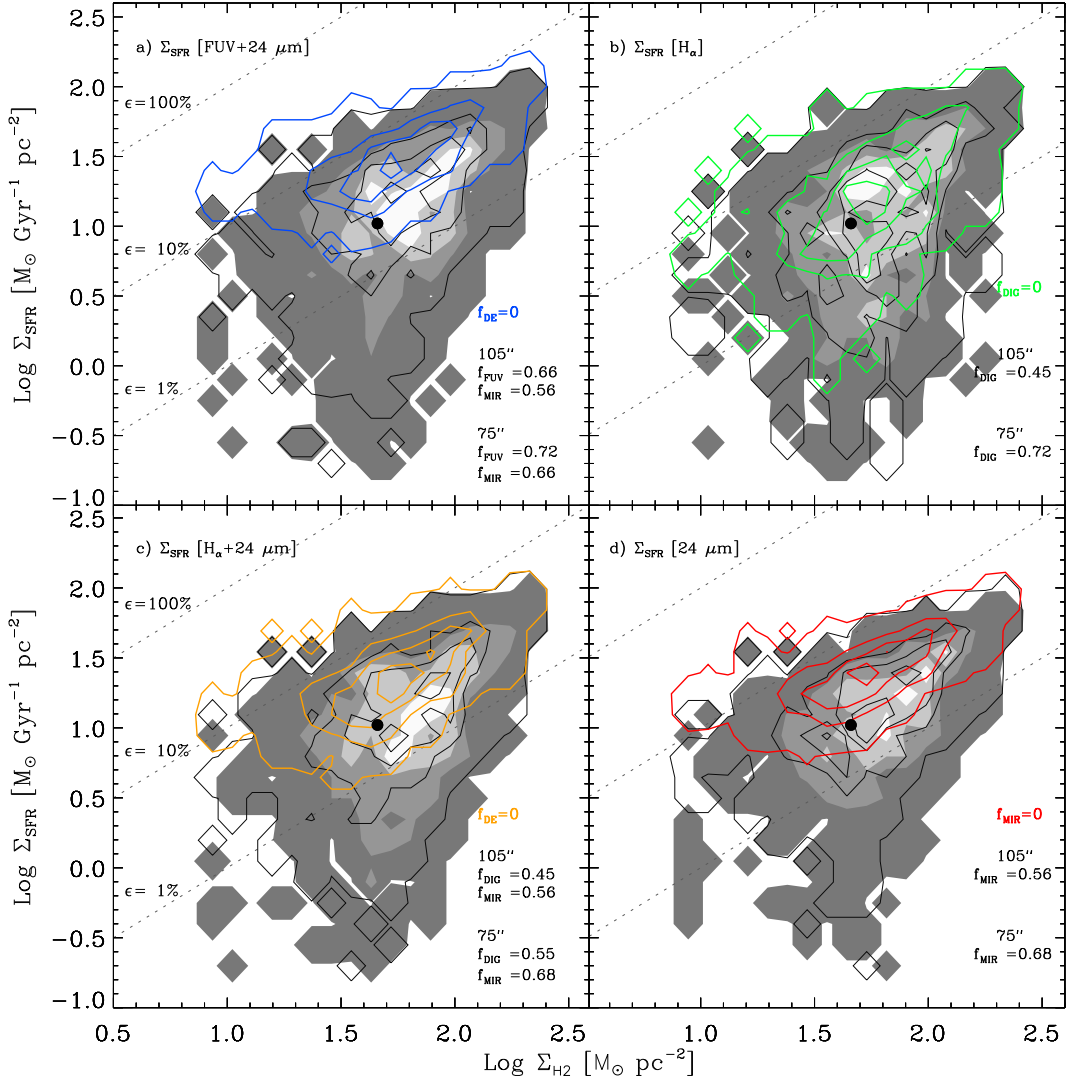}
\caption{Pixel-by-pixel analysis all four SFR tracers. The figure shows 
the molecular gas-SFR surface density relations at two representative 
filter scales, 105\arcsec\ (filled contours) and 180\arcsec\ (black
unfilled contours) as well as the case of no filtering, i.e., when 
\fde=0\ (colored unfilled contours). At any smoothing scale the amount 
of DE varies in the maps as shown at the bottom right in each panel 
(see also Table 2). The gray scale represent the two dimensional 
histogram of the frequency of points, and the contours are placed at 
90\%, 75\%, 50\%, and 25\% of the maximum frequency.
The diagonal dotted lines represent 
constant SFE ($\rm \epsilon$) where $\rm {\epsilon} \sim$100\%, 10\%, 
and 1\% correspond to gas depletion time scale of 0.1, 1.0, and 10 
Gyr. The filled circle represents the disk averaged estimate within 
the \optrad. The correlation strength varies among the maps. For a 
given SFR map, it depends on the degree of unsharp masking. The 
Pearson correlation coefficients at 105\arcsec\ and 180\arcsec\ are 
$\rm r\sim0.4-0.55$ (weak correlation) and $\sim0.55-0.7$ (strong 
correlation), respectively, indicating how the subtraction of DE 
influence the correlation between the points. 
A comparison of the distributions of points shown by filled and 
unfilled contours clearly shows that the subtraction of DE affects 
the low surface density regions. 
\label{pixdist}} 
\end{figure}

\begin{figure}[t]
\plotone{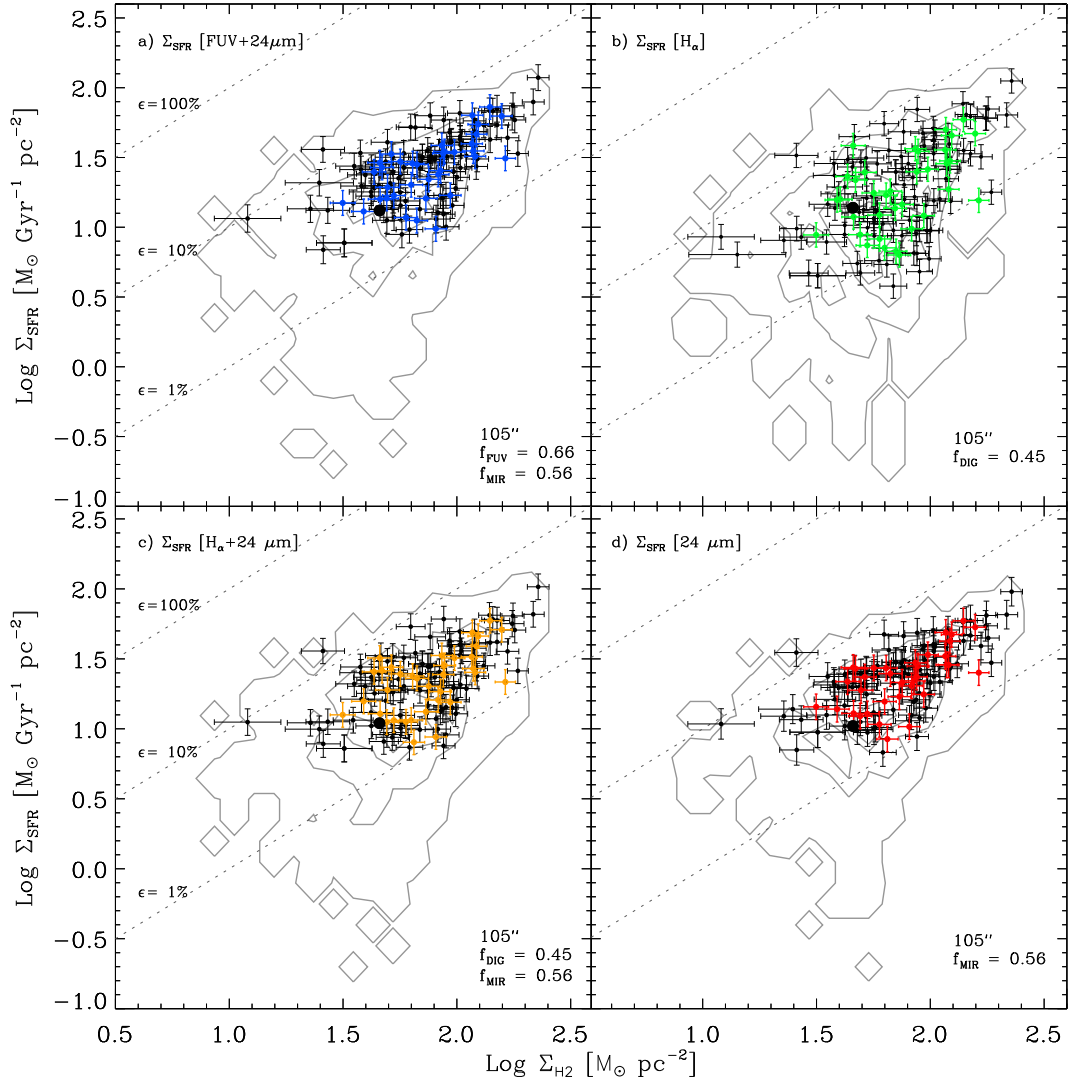}
\caption{Aperture analysis for all SFR tracers for the selected apertures 
as shown in Fig. \ref{apt}. The faint lines correspond to the contours 
of pixel distributions at 105\arcsec\ filter scale (shown as filled 
contours in Fig. \ref{pixdist}). The diagonal dotted lines present 
constant efficiencies as in Fig. \ref{pixdist}.
\label{aptdist}}
\end{figure}

\begin{figure}[t]
\epsscale{1.00}
\plotone{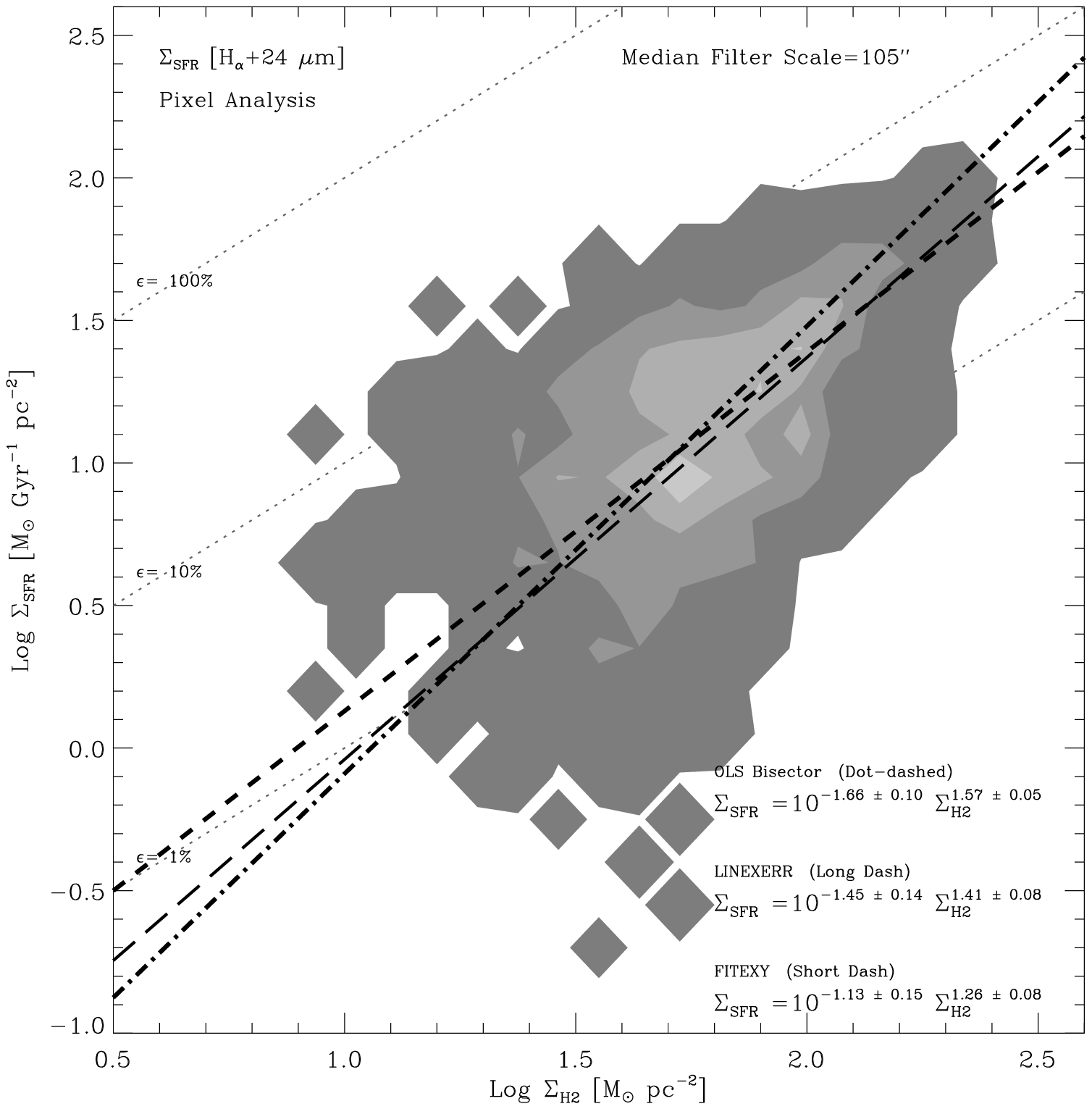}
\caption{Measurements provided by bivariate regression methods for 
\Scom\ - \Shtwo\ relation using pixel-by-pixel sampling. The SFR 
tracer is subject to unsharp making with a smoothing scale of 
105\arcsec\ which results in \fdig=0.45\ and \fmir=0.56. 
The gray scale represent the two dimensional histogram of the 
frequency of points, and the contours are placed at 90\%, 75\%, 50\%, 
and 25\% of the maximum frequency.
For this particular distribution of points, the 
FITEXY provides the shallowest slope ($\nmol \sim 1.3$ short dash line) 
The OLS bisector ($\nmol \sim 1.6$ dashed-dot line) and the LINEXERR 
($\nmol \sim 1.4$ long dash line) methods, on the other hand, provide 
steeper estimates. The estimated slope, zero point, and formal error 
provided by individual fit are shown at the lower right corner. The 
intrinsic scatter provided by FITEXY, OLS bisector, and LINEXERR 
methods are $\sigsct \sim 0.4$ 0.3, and 0.3 dex, respectively. 
The vertical and horizontal axes range from $\sim0.1-114 \ \msunyrpc$ 
and $\sim10-245 \ \msunpc$, respectively.
\label{pixfit}}
\end{figure}

\begin{figure}[t]
\epsscale{1.00}
\plotone{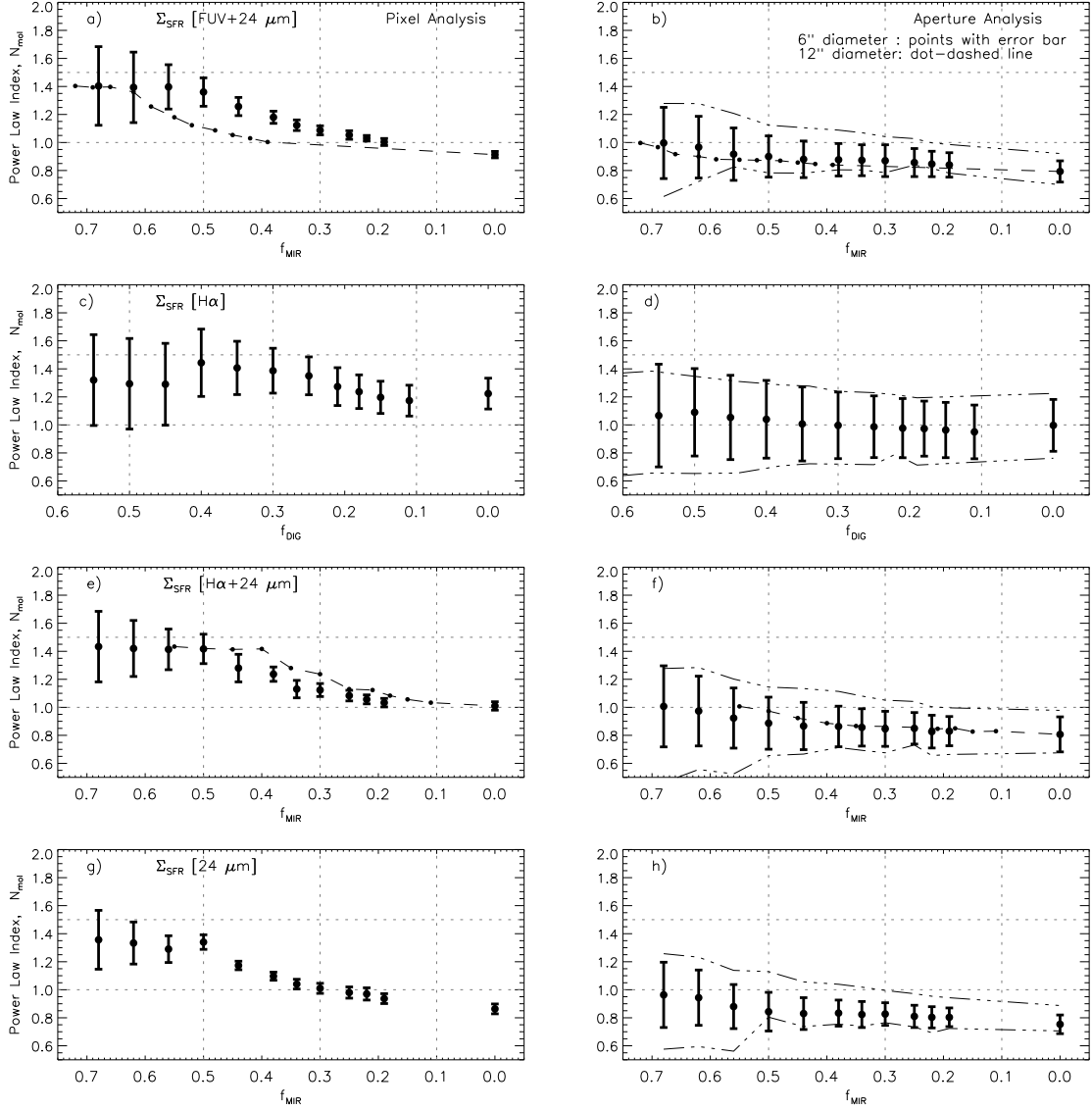}
\caption{Dependence of the power law index (\nmol) of the molecular 
gas SF law on various diffuse fractions. The figure summarizes results 
for both pixel (left panels) and aperture (right panels) samplings. 
Results from $6\arcsec \sim 500$ pc  (points) and $12\arcsec \sim 1$ 
kpc (dashed-dot line) diameter aperture samplings are shown in the 
right panels.
The methodological mean (\nmolfit) and dispersion (\sigfit) from 
three measurements are shown, respectively, by gray filled circles 
and vertical lines. 
The black points connected by dashed lines in panels a and b (e and f) 
represent the locus of \nmolfit\ at \ffuv (\fdig).
The horizontal dotted lines represent $\nmol=1$ and $\nmol=1.5$.
The vertical dotted lines demarcate the regions where the DE is assumed 
to be (from left to right) the dominant, significant, sub-dominant, and 
negligible component of the total disk emission. Each point along the 
horizontal axis has 15\% uncertainty (see appendix \ref{appen:mask} for 
details).  
\label{ind}}
\end{figure}

\begin{figure}[t]
\epsscale{1.00}
\plotone{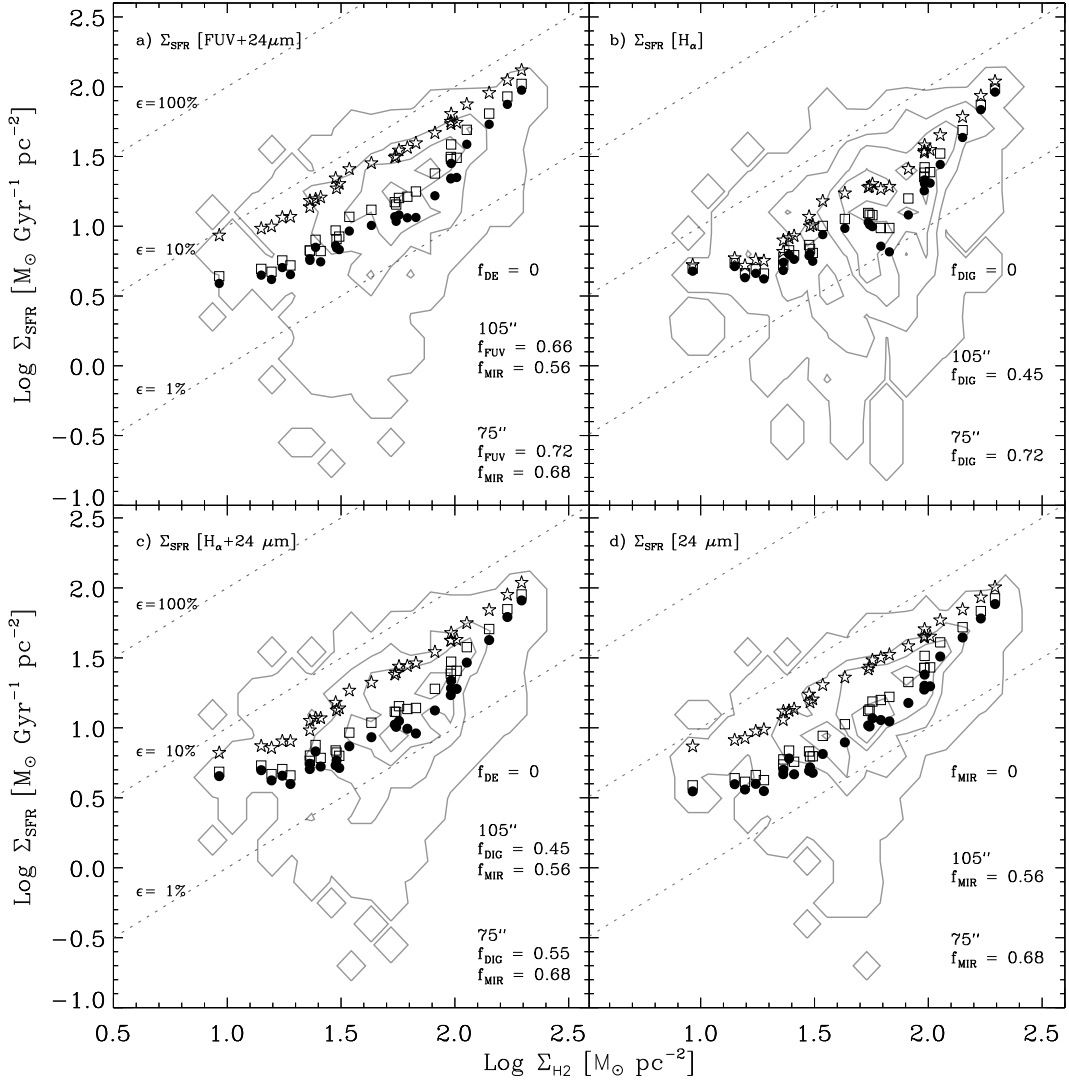}
\caption{Azimuthally averaged radial profile analysis all four SFR tracers. 
Figure shows the radial relations at the dominant (filled circle), 
significant (open square) and negligible ($\fde=0$) diffuse fractions. 
The faint lines correspond to the contours of pixel distributions at 
105\arcsec\ filter scale (shown as filled contours in Fig. \ref{pixdist}). 
The diagonal dotted lines represent constant efficiencies as in 
Fig. \ref{pixdist}. The bin size is 500 pc. The absence of an extended 
tail along the vertical axis at the low surface density regions, 
irrespective of the subtraction of the amount of diffuse fractions, 
indicates that azimuthally averaged radial profile systematically sample 
the high SFR surface density regions compared to pixel sampling. 
See Table \ref{radpro_table} for the derived fitted parameters for all 
SFR tracer maps at these three diffuse fractions. \label{radial}}
\end{figure}

\begin{figure}[t]
\epsscale{1.00}
\plotone{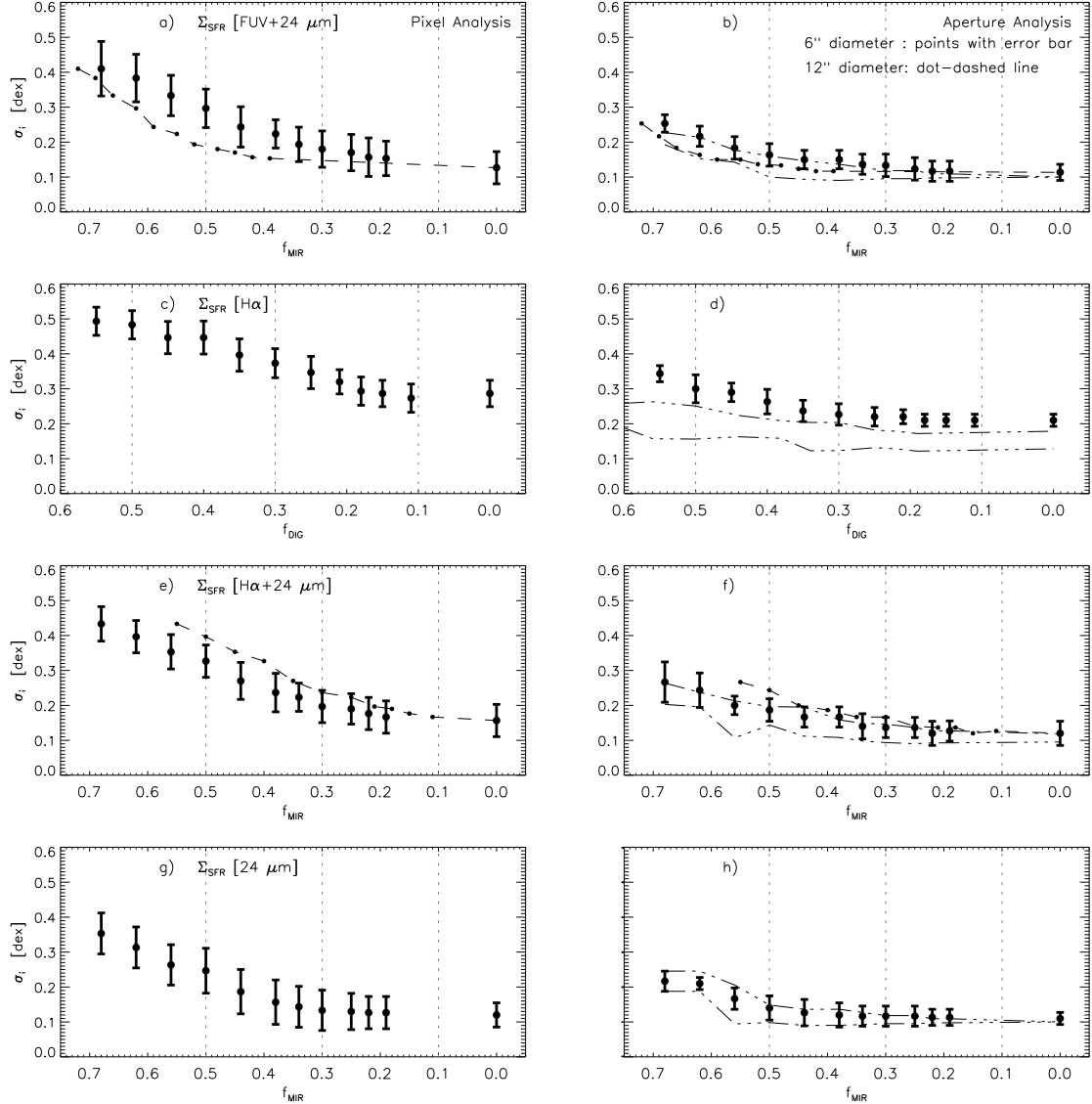}
\caption{The intrinsic scatter (\sigsct) in the gas-SFR surface density 
relations at various diffuse fractions for pixel (left panels) and 
aperture (right panels) samplings. 
Results from $6\arcsec \sim 500$ pc  (points) and $12\arcsec \sim 1$ 
kpc (dashed-dot line) diameter aperture samplings are shown in the 
right panels.
Figure shows the methodological mean of the observed scatter 
(\sigsctfit; ``mean dispersion'')by filled circle and the dispersion 
in the observed scatter (``dispersion of dispersion'') from linear 
regression methods by vertical solid lines.
Figure highlights that the estimate of intrinsic scatter depends on the 
choice of the SFR tracer as well as on the treatment of DE. 
For example, the scatter is systematically larger in the \Shtwo - \Shal\ 
relation (methodological dispersion, $\sigsctfit \sim 0.3-0.6$ dex). 
The scatter becomes smaller for larger aperture (dashed-dot line in the 
right panel) because of the reduction in the sampled points. The vertical 
lines have similar meanings as in Fig. \ref{ind}.
Each point along the horizontal axis has 15\% uncertainty (see appendix 
\ref{appen:mask} for details).   
\label{scatter}}
\end{figure}

\begin{figure}[t]
\epsscale{1.00}
\plotone{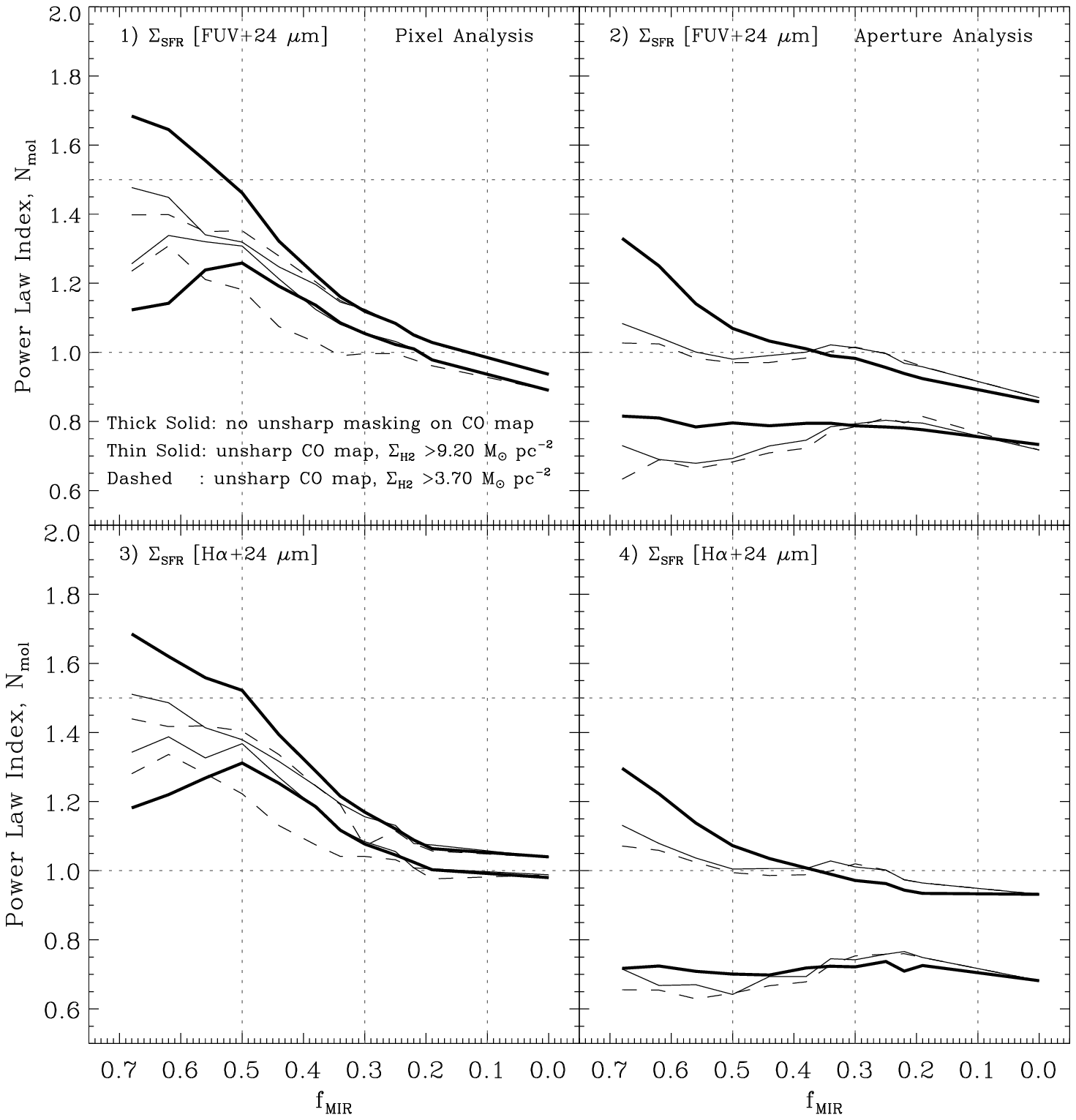}
\caption{Dependence of the power law index (\nmol) of molecular gas SF 
law when both SFR tracer and \cone\ map are subject to unsharp masking. 
The dashed and thin solid lines, respectively, show the methodological
dispersions when the selection threshold for the data in the \Shtwo\
unsharp-masked map is placed at 1$\sigma$ (3.7 \msunpc) and
2.5$\sigma$ (9.2 \msunpc).  The thick solid line illustrates the 
case when only the SFR tracers are subject to masking. Presentation 
style is similar to Fig. \ref{ind}.
\label{diffco}}
\end{figure}

\begin{figure}[t]
\epsscale{0.90}
\plotone{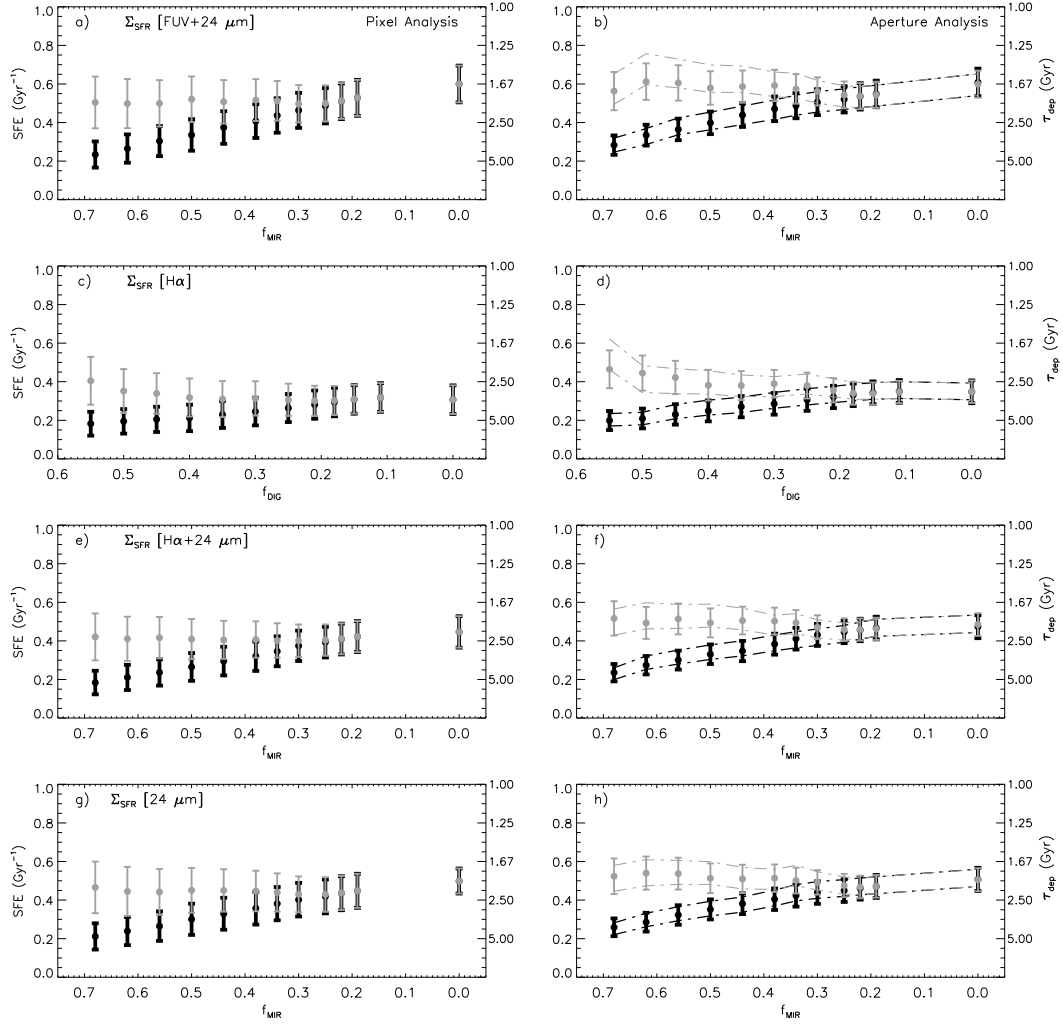}
\caption{Disk averaged SFE (Gyr$^{-1}$) and the depletion timescale, 
\tdep\ (Gyr), at various diffuse fractions. The SFE and the \tdep\ 
are shown on the left and right vertical axis, respectively. Results
for both (6\arcsec\ by points and 12\arcsec\ by dashed-dot line)
aperture samplings are shown in the right panels.  Parameters
estimated from applying unsharp masking to the SFR tracer only, 
and to both the SFR and the gas maps, are shown in black and gray
respectively.  For a given SFR tracer, the subtraction of DE
introduces $\sim$50\% (\Shal) up to a factor 3 variations in the
global SFE. At a given \fde, the global SFE derived from four 
tracers are approximately consistent with one another (see Table
\ref{dap_table}), although \halpha\ yields a lower depletion time. 
The dashed lines added by black points shown in Figs. \ref{ind} and
\ref{scatter} have omitted for clarity of presentation. 
The vertical lines have similar meanings as in Fig. \ref{ind}. 
\label{sfede}}
\end{figure}


\tabletypesize{\scriptsize} 
\begin{deluxetable*}{@{}c@{}|@{}c@{}|@{}ccc|@{}ccc|@{}ccc|@{}ccc}
\tablecaption{Fitted Parameters from Pixel Analysis}
\tablewidth{0pt}
\tablecolumns{14}
\tablehead{
\colhead{Filt.}
&\colhead{\fdig}   
&\colhead{} 
&\colhead{FUV+24 \um} 
&\colhead{} 
&\colhead{} 
&\colhead{\halpha} 
&\colhead{}
&\colhead{} 
&\colhead{\halpha+24 \um}
&\colhead{} 
&\colhead{}
&\colhead{24 \um} \\
\cline{3-14}
\colhead{($\arcsec$)}
&\colhead{}
&\colhead{\rm Log A}
&\colhead{\nmol}  
&\colhead{\sigsct}  
&\colhead{\rm Log A}  
&\colhead{\nmol}
&\colhead{\sigsct}  
&\colhead{\rm Log A}  
&\colhead{\nmol}
&\colhead{\sigsct}  
&\colhead{\rm Log A}  
&\colhead{\nmol}  
&\colhead{\sigsct}       
} 
\startdata 
&
&${-1.97 \pm 0.11}$ &$1.71 \pm 0.06$ &0.50
&${-1.99 \pm 0.13}$ &$1.67 \pm 0.07$ &0.54
&${-2.04 \pm 0.11}$ &$1.70 \pm 0.06$ &0.49
&${-1.75 \pm 0.10}$ &$1.57 \pm 0.05$ &0.42 \\

75 &0.55
&${-0.96 \pm 0.14}$ &$1.15 \pm 0.07$ &0.37
&${-0.83 \pm 0.20}$ &$1.02 \pm 0.11$ &0.47
&${-1.12 \pm 0.17}$ &$1.19 \pm 0.09$ &0.40
&${-1.00 \pm 0.13}$ &$1.15 \pm 0.07$ &0.33 \\

&
&${-1.36 \pm 0.16}$ &$1.35 \pm 0.09$ &0.36
&${-1.29 \pm 0.21}$ &$1.25 \pm 0.11$ &0.47
&${-1.55 \pm 0.18}$ &$1.40 \pm 0.10$ &0.41
&${-1.40 \pm 0.15}$ &$1.35 \pm 0.08$ &0.31 \\ \cline{1-14} \\

&
&${-1.57 \pm 0.10}$ &$1.57 \pm 0.05$ &0.40
&${-1.79 \pm 0.10}$ &$1.60 \pm 0.05$ &0.50
&${-1.66 \pm 0.09}$ &$1.57 \pm 0.05$ &0.41
&${-1.27 \pm 0.08}$ &$1.38 \pm 0.04$ &0.33 \\
 
105 &0.45
&${-1.06 \pm 0.13}$ &$1.27 \pm 0.07$ &0.30
&${-0.77 \pm 0.16}$ &$1.02 \pm 0.09$ &0.42
&${-1.13 \pm 0.15}$ &$1.26 \pm 0.08$ &0.32
&${-0.95 \pm 0.12}$ &$1.19 \pm 0.06$ &0.23 \\

&
&${-1.26 \pm 0.13}$ &$1.36 \pm 0.07$ &0.30
&${-1.22 \pm 0.18}$ &$1.25 \pm 0.10$ &0.42
&${-1.45 \pm 0.14}$ &$1.41 \pm 0.08$ &0.33
&${-1.19 \pm 0.11}$ &$1.30 \pm 0.06$ &0.23 \\ \cline{1-14} \\

&
&${-0.50 \pm 0.05}$ &$1.10 \pm 0.03$ &0.24
&${-1.28 \pm 0.07}$ &$1.41 \pm 0.04$ &0.36
&${-0.68 \pm 0.05}$ &$1.15 \pm 0.05$ &0.25
&${-0.34 \pm 0.04}$ &$0.98 \pm 0.02$ &0.20 \\

180 &0.21
&${-0.44 \pm 0.09}$ &$1.05 \pm 0.05$ &0.15
&${-0.83 \pm 0.13}$ &$1.15 \pm 0.07$ &0.30
&${-0.57 \pm 0.09}$ &$1.07 \pm 0.05$ &0.17
&${-0.42 \pm 0.07}$ &$1.00 \pm 0.04$ &0.10 \\

&
&${-0.58 \pm 0.08}$ &$1.11 \pm 0.04$ &0.15
&${-1.10 \pm 0.12}$ &$1.27 \pm 0.07$ &0.30
&${-0.74 \pm 0.08}$ &$1.15 \pm 0.04$ &0.17
&${-0.53 \pm 0.07}$ &$1.05 \pm 0.04$ &0.10 \\ \cline{1-14} \\

&
&${-0.03 \pm 0.04}$ &$0.90 \pm 0.02$ &0.18
&${-1.10 \pm 0.06}$ &$1.33 \pm 0.03$ &0.33
&${-0.36 \pm 0.04}$ &$1.01 \pm 0.02$ &0.21
&${-0.02 \pm 0.04}$ &$0.82 \pm 0.02$ &0.16 \\

-- &0.0
&${-0.06 \pm 0.06}$ &$0.90 \pm 0.03$ &0.10
&${-0.74 \pm 0.11}$ &$1.12 \pm 0.06$ &0.27
&${-0.34 \pm 0.07}$ &$0.98 \pm 0.04$ &0.13
&${-0.07 \pm 0.05}$ &$0.86 \pm 0.03$ &0.10 \\

&
&${-0.16 \pm 0.06}$ &$0.94 \pm 0.03$ &0.10
&${-0.97 \pm 0.11}$ &$1.23 \pm 0.06$ &0.26
&${-0.47 \pm 0.07}$ &$1.04 \pm 0.04$ &0.13
&${-0.16 \pm 0.05}$ &$0.90 \pm 0.03$ &0.10
\enddata 
\tablecomments{The normalization constant, power-law index, 
and intrinsic scatter of the molecular gas SF law are derived 
from unsharp masked SFR tracers (i.e., unsharp masking along 
the vertical axis). The filter scales from the top correspond 
to the dominant ($\fdig \gtrsim 50\%$), significant 
($\fdig \sim 30-50\%$), and sub-dominant ($\fdig \sim 10-30\%$) 
diffuse fractions. The bottom row shown by the dashed line 
represents \fdig=0. At a given filter scale, the three rows 
under FUV+24\um\ provide the measurements of the OLS bisector, 
FITEXY, and LINEXERR fitting method, respectively. 
Typical uncertainty in the intrinsic scatter is few percent.} 
\label{pixel_table}
\end{deluxetable*}

\begin{deluxetable*}{@{}c@{}|@{}c@{}|@{}ccc|@{}ccc|@{}ccc|@{}ccc}
\tabletypesize{\scriptsize}
\tablecaption{Fitted Parameters from 6\arcsec\ Aperture Analysis}
\tablewidth{0pt}
\tablehead{
\colhead{Filt.}   
&\colhead{\fdig} 
&\colhead{} 
&\colhead{FUV+24 \um} 
&\colhead{} &\colhead{} 
&\colhead{\halpha} 
&\colhead{} &\colhead{} 
&\colhead{\halpha+24 \um} 
&\colhead{} &\colhead{} 
&\colhead{24 \um} \\
\cline{3-14}
\colhead{($\arcsec$)}
&\colhead{}
&\colhead{\rm Log A}
&\colhead{\nmol} 
&\colhead{\sigsct}  
&\colhead{\rm Log A}  
&\colhead{\nmol}
&\colhead{\sigsct}  
&\colhead{\rm Log A}  
&\colhead{\nmol}
&\colhead{\sigsct}  
&\colhead{\rm Log A}  
&\colhead{\nmol}    
&\colhead{\sigsct}     
} 
\startdata 
&
&${-0.09 \pm 0.17}$ &$1.29 \pm 0.09$ &0.28
&${-1.61 \pm 0.22}$ &$1.49 \pm 0.11$ &0.37
&${-1.27 \pm 0.17}$ &$1.34 \pm 0.08$ &0.30
&${-1.03 \pm 0.16}$ &$1.23 \pm 0.08$ &0.25 \\

75 &0.55
&${-0.26 \pm 0.23}$ &$0.84 \pm 0.12$ &0.25
&${-0.40 \pm 0.37}$ &$0.85 \pm 0.20$ &033
&${-0.30 \pm 0.23}$ &$0.82 \pm 0.12$ &0.20
&${-0.24 \pm 0.23}$ &$0.80 \pm 0.12$ &0.20 \\

&
&${-0.30 \pm 0.26}$ &$0.86 \pm 0.14$ &0.23 
&${-0.51 \pm 0.35}$ &$0.90 \pm 0.19$ &0.33
&${-0.39 \pm 0.28}$ &$0.86 \pm 0.15$ &0.30
&${-0.35 \pm 0.24}$ &$0.86 \pm 0.13$ &0.20 \\ \cline{1-14} \\

& 
&${-0.67 \pm 0.14}$ &$1.13 \pm 0.07$ &0.22
&${-0.34 \pm 0.18}$ &$1.40 \pm 0.09$ &0.32
&${-0.84 \pm 0.14}$ &$1.17 \pm 0.07$ &0.23
&${-0.61 \pm 0.13}$ &$1.06 \pm 0.07$ &0.20 \\
 
105 &0.45
&${-0.03 \pm 0.16}$ &$0.78 \pm 0.08$ &0.16
&${-0.33 \pm 0.30}$ &$0.86 \pm 0.16$ &0.27
&${-0.10 \pm 0.21}$ &$0.78 \pm 0.11$ &0.18
&${-0.06 \pm 0.18}$ &$0.77 \pm 0.10$ &0.16 \\

&
&${-0.13 \pm 0.20}$ &$0.84 \pm 0.11$ &0.17
&${-0.42 \pm 0.30}$ &$0.90 \pm 0.16$ &0.28
&${-0.19 \pm 0.22}$ &$0.82 \pm 0.12$ &0.19
&${-0.14 \pm 0.19}$ &$0.81 \pm 0.10$ &0.14 \\ \cline{1-14} \\

&
&${-0.25 \pm 0.12}$ &$0.99 \pm 0.06$ &0.17
&${-0.87 \pm 0.14}$ &$1.22 \pm 0.07$ &0.24
&${-0.35 \pm 0.11}$ &$0.99 \pm 0.06$ &0.17
&${-0.20 \pm 0.10}$ &$0.92 \pm 0.05$ &0.15 \\

180  &0.21 
&${+0.12 \pm 0.16}$ &$0.79 \pm 0.08$ &0.12
&${-0.16 \pm 0.20}$ &$0.84 \pm 0.11$ &0.22
&${+0.07 \pm 0.15}$ &$0.76 \pm 0.08$ &0.12 
&${+0.06 \pm 0.15}$ &$0.78 \pm 0.08$ &0.10 \\

&
&${-0.04 \pm 0.16}$ &$0.82 \pm 0.08$ &0.11
&${-0.23 \pm 0.23}$ &$0.87 \pm 0.12$ &0.20
&${-0.02 \pm 0.17}$ &$0.79 \pm 0.09$ &0.12
&${-0.05 \pm 0.14}$ &$0.78 \pm 0.07$ &0.10 \\ \cline{1-14} \\

&
&${-0.03 \pm 0.10}$ &$0.88 \pm 0.05$ &0.14
&${-0.83 \pm 0.15}$ &$1.21 \pm 0.08$ &0.23
&${-0.21 \pm 0.11}$ &$0.95 \pm 0.06$ &0.16
&${+0.03 \pm 0.09}$ &$0.83 \pm 0.05$ &0.13 \\
 
-- &0.0
&${+0.29 \pm 0.12}$ &$0.74 \pm 0.07$ &0.10
&${-0.20 \pm 0.18}$ &$0.88 \pm 0.09$ &0.20
&${+0.16 \pm 0.16}$ &$0.75 \pm 0.08$ &0.10
&${+0.24 \pm 0.13}$ &$0.71 \pm 0.06$ &0.10 \\

&
&${+0.24 \pm 0.14}$ &$0.76 \pm 0.07$ &0.10
&${-0.25 \pm 0.22}$ &$0.90 \pm 0.12$ &0.20
&${-0.08 \pm 0.15}$ &$0.78 \pm 0.08$ &0.10
&${+0.22 \pm 0.12}$ &$0.72 \pm 0.06$ &0.10
\enddata 
\tablecomments{The normalization constant, power-law index, 
and intrinsic scatter of the molecular gas SF law are derived 
from unsharp masked SFR tracers (i.e., unsharp masking along 
the vertical axis).
The filter scales from the top correspond to the dominant 
($\fdig \gtrsim 50\%$), significant ($\fdig \sim 30-50\%$), and 
sub-dominant ($\fdig \sim 10-30\%$) diffuse fractions. The bottom 
row shown by the dashed line represents \fdig=0. 
At a given filter scale, the three rows under FUV+24\um\ provide 
the measurements of the OLS bisector, FITEXY, and LINEXERR fitting 
method, respectively. Typical uncertainty in the intrinsic scatter 
is few percent. }
\label{apt1_table}
\end{deluxetable*}

\begin{deluxetable*}{@{}c@{}|@{}c@{}|@{}ccc|@{}ccc|@{}ccc|@{}ccc}
\tabletypesize{\scriptsize}
\tablecaption{Fitted Parameters from 12\arcsec\ Aperture Analysis}
\tablewidth{0pt}
\tablehead{
\colhead{Filt.}   
&\colhead{\fdig} 
&\colhead{} 
&\colhead{FUV+24 \um} 
&\colhead{} &\colhead{} 
&\colhead{\halpha} 
&\colhead{} &\colhead{} 
&\colhead{\halpha+24 \um} 
&\colhead{} &\colhead{} 
&\colhead{24 \um} \\
\cline{3-14}
\colhead{($\arcsec$)}
&\colhead{}
&\colhead{\rm Log A}
&\colhead{\nmol}  
&\colhead{\sigsct}  
&\colhead{\rm Log A}  
&\colhead{\nmol}
&\colhead{\sigsct}  
&\colhead{\rm Log A}  
&\colhead{\nmol}
&\colhead{\sigsct}  
&\colhead{\rm Log A}  
&\colhead{\nmol}  
&\colhead{\sigsct}  
} 
\startdata
&
&${-1.20 \pm 0.30}$ &$1.33 \pm 0.15$ &0.23
&${-1.42 \pm 0.33}$ &$1.39 \pm 0.17$ &0.26
&${-1.30 \pm 0.33}$ &$1.34 \pm 0.16$ &0.26
&${-1.22 \pm 0.30}$ &$1.31 \pm 0.15$ &0.25 \\

75  &0.55
&${-0.12 \pm 0.44}$ &$0.76 \pm 0.23$ &0.20
&${-0.31 \pm 0.67}$ &$0.81 \pm 0.36$ &0.20
&${-0.05 \pm 0.52}$ &$0.63 \pm 0.28$ &0.20
&${-0.11 \pm 0.50}$ &$0.73 \pm 0.26$ &0.20 \\

&
&${-0.10 \pm 0.50}$ &$0.75 \pm 0.26$ &0.20
&${-0.14 \pm 0.56}$ &$0.73 \pm 0.29$ &0.25
&${+0.04 \pm 0.56}$ &$0.64 \pm 0.30$ &0.24
&${-0.07 \pm 0.49}$ &$0.71 \pm 0.26$ &0.20 \\ \cline{1-14} \\

&
&${-0.88 \pm 0.27}$ &$1.22 \pm 0.13$ &0.18
&${-1.45 \pm 0.35}$ &$1.44 \pm 0.18$ &0.25
&${-1.01 \pm 0.26}$ &$1.25 \pm 0.13$ &0.20
&${-0.84 \pm 0.25}$ &$1.17 \pm 0.12$ &0.21 \\
 
105 &0.45
&${-0.39 \pm 0.84}$ &$0.99 \pm 0.42$ &0.15
&${-0.21 \pm 0.85}$ &$0.82 \pm 0.43$ &0.15
&${+0.25 \pm 0.73}$ &$0.62 \pm 0.37$ &0.10
&${+0.30 \pm 0.42}$ &$0.61 \pm 0.21$ &0.10 \\

&
&${-0.15 \pm 0.39}$ &$0.84 \pm 0.21$ &0.15
&${-0.22 \pm 0.56}$ &$0.80 \pm 0.29$ &0.23
&${-0.02 \pm 0.44}$ &$0.72 \pm 0.23$ &0.18
&${-0.09 \pm 0.37}$ &$0.77 \pm 0.20$ &0.14 \\ \cline{1-14} \\

&
&${-0.41 \pm 0.21}$ &$1.06 \pm 0.11$ &0.12
&${-1.01 \pm 0.25}$ &$1.28 \pm 0.13$ &0.20
&${-0.54 \pm 0.21}$ &$1.08 \pm 0.11$ &0.15
&${-0.39 \pm 0.21}$ &$1.01 \pm 0.10$ &0.12 \\

180 &0.21
&${-0.03 \pm 0.22}$ &$0.82 \pm 0.12$ &0.10
&${-0.17 \pm 0.44}$ &$0.83 \pm 0.23$ &0.12
&${-0.09 \pm 0.27}$ &$0.75 \pm 0.14$ &0.10
&${-0.07 \pm 0.25}$ &$0.84 \pm 0.14$ &0.10 \\

&
&${-0.05 \pm 0.29}$ &$0.86 \pm 0.15$ &0.10
&${-0.16 \pm 0.42}$ &$0.83 \pm 0.22$ &0.17
&${-0.06 \pm 0.32}$ &$0.76 \pm 0.17$ &0.11
&${+0.02 \pm 0.27}$ &$0.79 \pm 0.14$ &0.10 \\ \cline{1-14} \\

&
&${-0.09 \pm 0.18}$ &$0.93 \pm 0.09$ &0.10
&${-0.94 \pm 0.26}$ &$1.26 \pm 0.13$ &0.18
&${-0.34 \pm 0.20}$ &$1.00 \pm 0.10$ &0.12
&${-0.11 \pm 0.18}$ &$0.90 \pm 0.01$ &0.10 \\

-- &0.0
&${-0.31 \pm 0.31}$ &$0.71 \pm 0.17$ &0.10
&${-0.16 \pm 0.32}$ &$0.85 \pm 0.17$ &0.13
&${-0.19 \pm 0.30}$ &$0.72 \pm 0.16$ &0.10
&${-0.15 \pm 0.15}$ &$0.76 \pm 0.08$ &0.10 \\

&
&${+0.17 \pm 0.25}$ &$0.79 \pm 0.13$ &0.10
&${-0.20 \pm 0.39}$ &$0.87 \pm 0.20$ &0.15
&${-0.12 \pm 0.28}$ &$0.76 \pm 0.15$ &0.10
&${+0.21 \pm 0.24}$ &$0.73 \pm 0.13$ &0.10
\enddata 
\tablecomments{The normalization constant, power-law 
index, and intrinsic scatter of the molecular gas SF law 
are derived from unsharp masked SFR tracers (i.e., unsharp 
masking along the vertical axis).
The filter scales from the top correspond to the dominant 
($\fdig \gtrsim 50\%$), significant ($\fdig \sim 30-50\%$), 
and sub-dominant ($\fdig \sim 10-30\%$) diffuse fractions. 
The bottom row shown by the dashed line represents \fdig=0. 
At a given filter scale, the three rows under FUV+24\um\ 
provide the measurements of the OLS bisector, FITEXY, and 
LINEXERR fitting method, respectively. Typical uncertainty 
in the intrinsic scatter is few percent.}
\label{apt2_table}
\end{deluxetable*}

\begin{deluxetable}{cc|cc|cc|cc|cc}
\tabletypesize{\scriptsize}
\tablecaption{Fitted Parameters from Azimuthally Averaged Radial 
Profile Analysis}
\tablewidth{0pt}
\tablehead{
\colhead{Filt.}   
&\colhead{\fdig} 
&\colhead{FUV+24 \um} 
&\colhead{} 
&\colhead{\halpha} 
&\colhead{} 
&\colhead{\halpha+24 \um} 
&\colhead{} 
&\colhead{24 \um} \\
\cline{3-10}
\colhead{($\arcsec$)}
&\colhead{}
&\colhead{\rm Log A}
&\colhead{\nmol}  
&\colhead{\rm Log A}  
&\colhead{\nmol}
&\colhead{\rm Log A}  
&\colhead{\nmol}
&\colhead{\rm Log A}  
&\colhead{\nmol}       
} 
\startdata 
75   &0.55
&${-0.82\pm0.11}$ &$1.13\pm0.07$
&${-0.81\pm0.15}$ &$1.09\pm0.09$
&${-0.79\pm0.13}$ &$1.07\pm0.08$
&${-0.94\pm0.11}$ &$1.16\pm0.06$ \\

105  &0.45
&${-0.78\pm0.08}$ &$1.17\pm0.05$
&${-0.78\pm0.12}$ &$1.11\pm0.07$
&${-0.76\pm0.10}$ &$1.11\pm0.06$
&${-0.87\pm0.07}$ &$1.18\pm0.04$ \\

180  &0.21
&${-0.53\pm0.04}$ &$1.12\pm0.02$
&${-0.63\pm0.08}$ &$1.09\pm0.05$
&${-0.55\pm0.05}$ &$1.08\pm0.03$
&${-0.57\pm0.03}$ &$1.10\pm0.02$ \\

--   &0.0
&${-0.16\pm0.03}$ &$0.98\pm0.02$
&${-0.64\pm0.06}$ &$1.12\pm0.04$
&${-0.37\pm0.04}$ &$1.02\pm0.02$
&${-0.20\pm0.02}$ &$0.95\pm0.01$ \\ \\ \cline{1-10} \\

75   &0.55
&${-0.16\pm0.03}$ &$0.98\pm0.02$
&${-0.64\pm0.06}$ &$1.12\pm0.04$
&${-0.37\pm0.04}$ &$1.02\pm0.02$
&${-0.20\pm0.02}$ &$0.95\pm0.01$ \\

105  &0.45
&${-0.35\pm0.06}$ &$1.07\pm0.04$
&${-0.38\pm0.08}$ &$1.04\pm0.05$
&${-0.35\pm0.06}$ &$1.02\pm0.04$
&${-0.45\pm0.05}$ &$1.09\pm0.03$ \\

180  &0.21
&${-0.20\pm0.10}$ &$1.08\pm0.06$
&${-0.33\pm0.08}$ &$1.07\pm0.05$
&${-0.24\pm0.08}$ &$1.05\pm0.06$
&${-0.24\pm0.10}$ &$1.06\pm0.04$ \\

--   &0.0 
&${-0.44\pm0.08}$ &$1.12\pm0.04$
&${-0.48\pm0.08}$ &$1.08\pm0.05$
&${-0.46\pm0.06}$ &$1.08\pm0.04$
&${-0.53\pm0.07}$ &$1.14\pm0.04$
\enddata 
\tablecomments{Top: parameters in the top four rows are derived 
from unsharp masked SFR tracer maps. Bottom: parameters in the 
bottom four rows are derived when both molecular gas and SFR maps 
are subject to unsharp masking. The intrinsic scatter \sigsct\ is 
0 in radial profile analysis and hence the parameters are obtained 
from the OLS bisector method. The filter scales from the top 
correspond to the dominant ($\fdig \gtrsim 50\%$), significant 
($\fdig \sim 30-50\%$), and sub-dominant (($\fdig \sim 10-30\%$) 
diffuse fractions. The bottom row shown by the dashed line 
represents \fdig=0.}
\label{radpro_table}
\end{deluxetable}

\clearpage

\appendix

\section{NGC~4254} 
\label{appen:ngc4254}

The optical structure of NGC~4254 shows a one-armed appearance
($\rm m=1$ mode), unlike most grand-design spirals where symmetric 
modes are prominent.  This unusual morphology has made it the target 
of several observational and numerical studies (Phookun \etal\ 1993;
Chemin \etal\ 2006; Sofue \etal\ 2003).  Various explanations for its
asymmetry have been put forth including the superposition of spiral
modes induced by global gravitational instability (Iye \etal\ 1982),
the asymmetric accretion of gas onto the disk (Phookun \etal\ 1993;
Bournaud \etal\ 2005), ram pressure stripping (Sofue \etal\ 2003;
Kantharia \etal\ 2008), a close high-speed encounter plus ram pressure
(Vollmer \etal\ 2005), a high-speed encounter only (Duc \& Bournaud 
2008), and harassment while entering Virgo (Haynes \etal\ 2007). On 
the other hand, detailed photometric studies of NGC~4254 show that 
it is photometrically similar to other Sc type spirals, and it has 
no close companion (Phookun \etal\ 1993).

NGC~4254 has a flat rotation curve with $v_{rot} \sim 150$ \kmsec\ 
up to 200\arcsec\ ($\sim 16$ kpc) from its center (Guhathakurta \etal\
1988). Apart from its morphological peculiarity, it does not show any
anomaly in its properties. For example, the distribution of SF 
through out the disk based on \halpha\ emission has been classified 
as normal (Koopman \& Kenney 2004). 
Its molecular gas fraction in the disk is similar to field galaxies, 
suggesting that it is entering Virgo for the first time and external 
agents (such as ram pressure) have not yet been effective at striping 
its gas, despite its morphology (Nakasihsi \etal\ 2006). NGC~4254 is 
a metal rich galaxy with a median $\rm 12 + \log[O/H] \sim9.0$ across 
the disk (Vila-Costas \& Edmunds 1992; Zaritsky \etal\ 1994).

\section {SFR Surface Density Maps}
\label{appen:sfrmap}

We construct four different SFR maps from combining FUV and MIR maps 
(Leroy \etal\ 2008), optical \halpha\ emission map (Kennicutt \etal\ 
2007; Prescott \etal\ 2007),  optical \halpha\ emission and MIR maps 
(Calzetti \etal\ 2007), and MIR map (Calzetti et al. 2007). The 
details of the construction of these SFR tracers can be found in the 
references. These different SFR tracers probe different time scales 
and hence the SF history of any particular galaxy. For example, the 
\halpha\ emission traces gas ionized by massive ($\rm M > 10 \ \msun$) 
stars over a timescale of $< 20$ Myr. The FUV luminosity corresponds 
relatively older ($< 100$ Myr), less massive ($\rm M \gtrsim 5 \ \msun$) 
stellar populations. The 
MIR 24 \um\ emission mostly traces re-processed radiation of newborn 
(few Myr) OB stellar associations embedded inside the parent molecular 
clouds. Although star clusters break from their parent clouds in less 
than 1 Myr, they remain associated with it for a much longer time scale, 
$t\sim10-30$ Myr. Thus this is the time scale associated with the 
MIR 24 \um\ emission as a tracer. 
However, if there is significant heating of the small dust 
grains by late B and A stars, the time scales associated with 24 \um\ 
emission may be as long as $\sim100$ Myr. For composite SFR tracer such 
as FUV + 24 \um\ or \halpha\ + 24 \um\, while the MIR emission traces the 
dust-obscured fraction of the SF, and the un-obscured SF can be traced 
either by FUV continuum or optical nebular emission.

Each pixel in the FUV, \halpha, and 24 \um\ emission maps has some 
uncertainties coming from observation and data processing. These errors 
propagate to the  uncertainties in the surface density maps. For example, 
uncertainties in the flux calibration, the stellar continuum subtraction, 
and the background subtraction contribute to the total error budget of 
\halpha. We assume 10\% uncertainty for each of these components and 
construct \halpha\ error map \erhal\ by adding these terms quadratically. 
A flat 10\% uncertainty in stellar continuum subtraction for the entire 
galaxy disk is probably unrealistic because stellar emission dominates in 
the galaxy center. Therefore, it will have higher contribution at the high 
surface brightness regions. We take this limit on ad hoc basis. The assumed 
fraction in the continuum subtraction may a lower limit since it varies 
along the galaxy disk contributing as much as 30\% to \imhal\ flux 
uncertainty (Koopman \etal\ 2001). Finally, there is a considerable 
(calibration) uncertainty in \halpha\ to SFR conversion (Calzetti \etal\ 
2007). We take all these factors into consideration when constructing the 
SFR error map (\erhal) from the H$\alpha$ image. 
The FUV and mid-IR error maps (\erfuv\ and \ermir) are constructed in a 
similar manner but without the contribution from the stellar continuum. 
The error in the composite SFR maps (FUV + 24 \um\ or \halpha\ + 24 \um) 
are constructed by combining the appropriate terms from the respective 
images.

\section{Molecular Gas Surface Density Map}
\label{appen:gasmap}

A conversion factor ($\cf$) is frequently used to determine the 
distribution of molecular hydrogen ($\htwo$) from the $\cone$ images.
In this study, we use $\cf=2.0 \times 10^{20}$ to be consistent with 
several other current studies (Wong \& Blitz 2002; Komugi \etal\ 2005; 
Gardan \etal\ 2007; Bigiel \etal\ 2008; Leroy \etal\ 2008). The choice 
of the calibration factor linearly scales the estimated gas densities.
It is worthwhile asking how accurate is the assumption of a single
conversion factor. Although the CO-to-H$_2$ appears to be
approximately constant for resolved GMCs (Bolatto \etal\ 2008), on the
hundreds of parsecs scales sampled in this study it may vary across
the disk, especially in the low density low metallicity regions (e.g.,
Garcia-Burillo \etal\ 1993).  We adopt a constant $\cf$ for simplicity
and consistency with most recent studies (Wong \& Blitz 2002; Komugi 
\etal\ 2005; Gardan \etal\ 2007; Bigiel \etal\ 2008; Leroy \etal\ 2008). 
Furthermore, we expect the effects of a varying conversion factor to 
be most important in the outer portions of the disk, which are beyond 
the central region sampled by this study. The \htwo\ surface densities 
are multiplied by a factor of 1.40 to account for the mass contribution 
of helium and are expressed in units of \msunpc. 

To obtain the integrated (interferometer + single) CO map, first, the 
CARMA cube was de-convolved using the implementation of the CLEAN 
algorithm in the MIRIAD task mossdi and the velocity channels (2.6 
\kmsec) in the IRAM cube has been re-binned to have the same velocity 
resolution (10 \kmsec) of the CARMA cube. The MIRIAD task immerge 
was then used to combine the CARMA and the IRAM cube in the image 
plane (e.g., Stanimirovi\'c \etal\ 1999). 
The integrated intensity map was created by convolving each plane of
the original data-cube with a 15\arcsec\ Gaussian to degrade its
resolution, selecting regions in each velocity plane where the signal
is larger than 2.5$\sigma$, and using these regions as masks on the
original cube to compute an integrated intensity. The corresponding
error map was computed by considering the $1\sigma$ rms value of 22
m\JBkmsec\ for each plane and multiplying by the square root of the
number of planes that make up each integrated intensity point. 
Although the RMS in each plane is approximately constant within 
the field-of-view of the mosaic, the noise in the integrated intensity 
map is a function of position because of the change in line width of 
the CO emission.

\section{Data Sampling}
\label{appen:sample} 

Existing studies of spatially resolved SF law in galaxies use three 
different data sampling methodologies. These are: azimuthally averaged 
radial profile (Kennicutt 1989; Martin \& Kennicutt 2001; Wong 
\& Blitz 2002; Boissier \etal\ 2003; Heyer \etal\ 2004; Schuster \etal\ 
2007; Bigiel \etal\ 2008); aperture analysis encompassing the star 
forming regions and centering on \halpha\ and MIR 24 \um\ emission peaks 
(Kennicutt \etal\ 2007; Blanc \etal\ 2009); and finally, pixel-by-pixel 
analysis (Bigiel \etal\ 2008; Leroy \etal\ 2008). 
To understand the methodological impact on the determination of the local 
SF law we have incorporated all three methods in our study  since each of 
these methodologies has strengths and weaknesses. 

Pixel analysis probes the gas-SFR density relation at the smallest
spatial resolution.  Aperture averages, on the other hand, provide
information relevant to ensemble averages of representative
regions. Azimuthally averaged radial profiles, in general, provide
estimates over much larger scales. This method of sampling produces
overall radial trends suppressing local variations. Radial profiles,
by construction, are less susceptible to angular resolution.  Although
the first sampling probes the smallest spatial scales in the map, it
is heavily dominated by its low surface brightness regions. The latter
two samplings, by construction, probe only the high surface brightness
regions where our signal-to-noise is best and DE in both the SFR and
the molecular gas tracers is likely to be a minor contaminant.

In our pixel analysis, we consider regions of the galaxy where both
\Ssfr\ and \Shtwo\ are higher than $2.5\sigma$. All images are
resolution-matched to 6\arcsec, and re-sampled at the Nyquist rate
($3\arcsec\times3\arcsec$ pixels). For the aperture analysis we lay
down a total of 102 circular apertures 6\arcsec\ (500 pc) in diameter,
in regions of high surface density, mostly following the spiral arms
(see Fig. \ref{apt}). These apertures are approximately independent,
with only a small fraction (5\%) of overlap. We also consider 34
circular apertures 12\arcsec\ (1 kpc) in diameter with a similar
overlapping fraction. It should be stressed here that sampling the
same region in all four SFR surface density maps simultaneously
determines the total number of apertures. The aperture centers for
both small and large radii were chosen independently. For any \Ssfr\
map all the apertures combined cover approximately $\approx45\%$ of
the total area covered by all the pixels above $2.5\sigma$
limit. Because of the nature of the selection method the aperture
samplings contain $\sim65\%$ of the total emission when compared pixel
sampling.

\section{Statistical Methodology and Fitting Procedures}
\label{appen:stat} 

The SF law, as expressed by Eq. 1, is a power law relationship between
the SFR surface density and the gas surface density.  The outcome of
any regression analysis, with the object of finding $\rm A$ and $\rm
N$, depends on the treatment of the data and its measurement errors,
and on the intrinsic scatter of the observables.  The intrinsic
scatter reflects the variations of local physical properties of 
star-forming regions (for example, evolutionary stages, stellar
populations, metallicity, obscuration, etc). The measurement errors,
on the other hand, depend entirely upon the observations and the
subsequent data reduction. The factors that contribute to the
measurement errors include flux calibration, continuum subtraction,
and background subtraction.

The linear regression methods can be divided into two broad classes 
depending on whether the intrinsic scatter or the measurement error 
dominate.  When seeking a linear relationship between two variables 
in a data set where both of them have small measurement errors but 
substantial yet unknown intrinsic scatter, the ordinary least square 
(OLS) bisector method provides one the best solutions, (Isobe \etal\ 
1990; Feigelson \& Babu 1992).  

When both the dependent and the independent variables are subject to 
measurement errors and intrinsic scatter of comparable magnitude, the 
regression analysis becomes more complex.  Several bivariate regression 
methods have been developed to deal with astronomical problems but each 
method has its own advantages and disadvantages (see Feigelson \& Babu 
1992 for a detailed account). The most widely used bivariate regression 
analysis is based on the least squares technique (FITEXY; Press \etal 
1992 and references therein). 
Akritas \& Bershady (1996) extended the OLS bisector method incorporating 
measurement error and the intrinsic scatter. This estimator is known as 
the bivariate correlated error and scatter (BCES) estimator. Kelly (2007) 
developed a bivariate estimator based on Bayesian statistics (LINEXERR).

The FITEXY and LINEXERR estimators differ in the underlying assumption 
of the nature of the true relationship between independent and dependent 
variables. Both the BCES and LINEXERR estimators assume that the data 
points are scattered around the true linear relationship. The FITEXY, on 
the other hand, assumes that there is no intrinsic scatter and the true 
points lie exactly on a straight line, providing a solution for data 
showing a perfect correlation. While the BCES and LINEXERR estimators 
incorporate the correlated measurement error, FITEXY does not account 
for it.
Correlated error arises when the dependent and the independent variables 
both are subjected to the same uncertainty. For example, surface densities 
are defined as the ratio of the total mass to the de-projected area of the 
disk. An uncertainty in the inclination measurement  leads to a correlated 
measurement error in the surface density of gas and SFR. The covariance 
term broadens the actual distribution of data points and thus provides a 
flatter relationship if unaccounted for (Akritas \& Bershady 1996). 

In this study we have used the OLS bisector, FITEXY, and LINEXERR 
estimators. We note here that while the OLS bisector method will guard 
our analysis against possible flaw in constructing measurement error maps, 
the other two methods will be required for a complete analysis of data.
Our experience shows that the BCES bisector method is highly sensitive to 
the measurement error resulting in unstable slope estimates compared to 
the three other methods mentioned above. We do not use this estimator in 
our analysis. 


We construct surface density measurement error maps as mentioned in 
section $\S$\ref{data}. The covariant term for each pixel is calculated 
from the error map. The intrinsic scatter is directly provided by the 
LINEXERR estimator. For the OLS bisector and FITEXY estimators we estimate 
it using the best fit line. To include scatter in the FITEXY estimator we 
iteratively adjust the error along the Y-axis iteratively until we achieve 
a reduced chi-squared $\sim 1$. For this estimator, the total dispersion 
along the y-axis is  $\rm \sigma_{y}^2 = \sigma_{m}^2 + \sigma_{i}^2$. 
The adjustment is made for $\sigsct$ which is a measure of the intrinsic 
dispersion in the gas-SFR surface density relation, while $\sigma_m$ is 
the measurement uncertainty obtained from error propagation. The formal 
error in each parameter quoted in this paper comes from the bootstrap 
sampling of 1000 realizations of the data points. 

\section{Unsharp Masking}
\label{appen:mask}

We use unsharp masking to model and remove the local variations of DE 
from the maps. The choice of the size of the median filter kernel plays 
a vital role in selecting the amount of diffuse component in the total 
disk emission. 
As mentioned earlier, we explore a number of filter sizes in each SFR 
tracer, carrying out our analysis for each case (see Table \ref{df_table}). 
We should stress here that our intention is to use the simplest model for 
a reliable estimate of the diffuse fraction in the disk of NGC~4254 which 
can be used with a reasonable confidence to explore the main goal of this 
study. 

For a filter size of 75\arcsec\ the fraction of the total emission 
contained in the smooth \halpha\ map is $\fdig\sim60\%$, on the acceptable 
high-end of the DIG fractions observed in other galaxies (Kennicutt \etal\ 
1995; Ferguson \etal\ 1996). Thus, DE in \halpha\ is likely overestimated 
for kernel sizes below 75\arcsec\ (corresponding to physical scales $\sim6$ 
kpc). 
Using smaller kernels for unsharp masking results in over-subtracted maps 
containing undesirable artifacts (many negative pixels and increased 
scatter).  The upper end of acceptable kernel sizes is determined by the 
image size. We set it roughly to two-third the image size, corresponding 
to $210\arcsec$ or a physical length of $\sim13$ kpc. With this scale we 
obtain $\fdig\sim15\%$ which is approximately the \fdig\ observed in the 
Galaxy (Reynolds 1991). Similar kernel sizes are also explored in the FUV 
and 24 \um\ maps. As expected, the diffuse fractions in the FUV, \halpha, 
and 24 \um\ emission maps vary substantially from one another. Within 
our range of kernel sizes, $\ffuv \sim 40-75\%$, while $\fmir\sim20-70\%$.

\begin{figure*}[t]
\epsscale{0.65}
\plotone{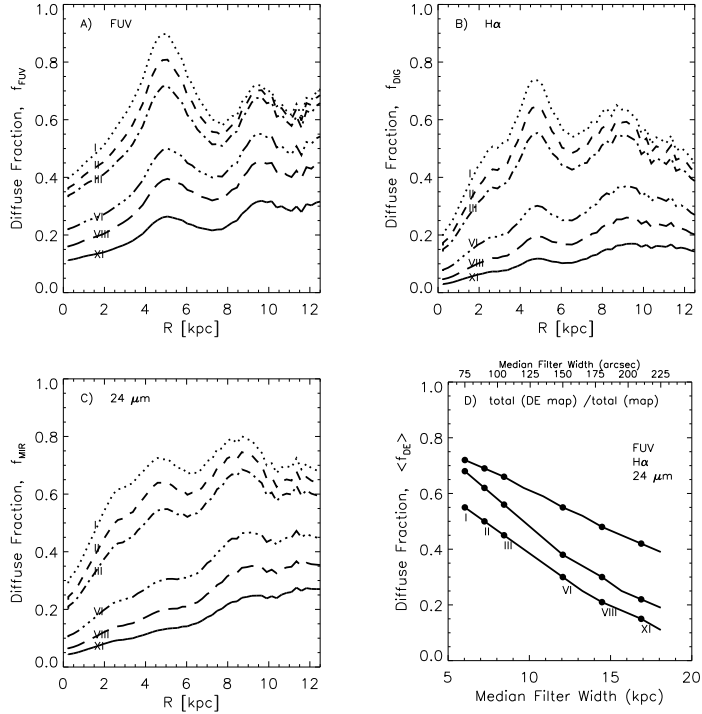}
\caption{Azimuthally averaged radial profiles of diffuse fractions. 
Panels A, B, and C show respectively, the \ffuv, \fdig, and \fmir. 
In each panel, lines with Roman numerals correspond to the DE maps 
constructed from six different filters as given in Table \ref{df_table}. 
Panel D shows the average diffuse fraction (${\rm{\bar f}_{DE}}$)
as a function of filter scale. 
The filled circles with roman numerals in panel D represent the radial 
averages of the corresponding lines in panels A, B, and C.
The figure shows clearly that \fde\ increases radially outward with a 
trend that is independent of wavelength or filter scale. The profiles 
flatten out at the edge of the disk (beyond $\sim$10 kpc).
For smaller smoothing scales both \fdig\ and \fmir\ vary as much as a 
factor of three along the disk. For longer smoothing 
scales the variation is about a factor of two for all \fde. The FUV 
map has the highest diffuse fraction compared to the other two maps 
(see Table 2).
Measurements for only six kernel sizes are shown for the ease of 
demonstration. The horizontal axes in panels A, B, and C span the 
optical radius, $\optrad \sim$12.5 kpc. 
\label{de}}
\end{figure*}

We show our azimuthally averaged radial distributions of DE fractions for 
NGC~4254 in Fig. \ref{de}. Panels A, B, and C show the radial profiles for 
\ffuv, \fdig, and \fmir\ respectively. In each panel, we show our results 
for six different smoothing filters.

It is evident from the figure that the \fde\ profiles increase radially 
outward. In spite of their distinct origins, all three diffuse fractions 
show this remarkably similar trend. Beyond $\sim$10 kpc these profiles 
flatten out. For smaller smoothing scales, \fmir\ varies as much as a 
factor of four whereas \fdig\ varies up to a factor three. The \ffuv\ 
shows a factor of two radial variation. For longer smoothing scales the 
variation is about a factor of two for all \fde. For \fdig, a similar 
trend was observed in the Sculptor group spiral NGC~7793 (Ferguson 
\etal\ 1996). 
Panel D shows the average fraction of diffuse emission, 
${\rm{\bar f}_{DE}}$, as a function of smoothing scale, summarizing 
the results of the first three panels.  The filled circles with roman 
numerals on the color coded line represent the radial average of the 
corresponding lines in panels A, B, and C. It is interesting to note 
that NGC~4254 is not unusual compared to the Local Group galaxies in 
terms of its multi-wavelength diffuse components.  As expected, the 
FUV map has the highest diffuse fraction compared to the MIR 24 \um\ 
and \halpha\ maps.  The \ffuv\ has slowest variation ($\sim40-75\%$) 
with the smoothing scale. The \fdig, on the other hand, falls sharply 
with larger smoothing scales. The amount of diffuse fraction in the 
24 \um\ map is intermediate between those in the \halpha\ and the FUV 
images.
 
At every filter scale, we combine the DE subtracted FUV, \halpha, and 24 
\um\ emission maps to construct the desired \Ssfr\ maps. Finally, a small 
global value is subtracted from the images. As long as most of the
pixels sample the background, sky background subtraction can be
considered an extreme case of unsharp masking where the median filter
size is the same as the entire image. In this case, DE estimated from
all three maps is negligibly small and this what we call as the zero
diffuse fraction ($\fde=0$) in the remaining sections. For each \Ssfr\
map as discussed in Appendix \ref{appen:sfrmap}, we use the \Shtwo\
map derived from the combination of single-dish and interferometer
data.

To remove the diffuse extended component from the CO distribution 
we employ the same techniques we have applied to the SFR tracers. 
To simplify the range of parameter space to explore we use filtering
kernels of the same size in both the \Ssfr\ and the \Shtwo\
maps. Although this is not necessarily correct, we take it as an
illustration of the effects of removing a diffuse
contribution in both axes.

\subsection{Image PSF and Unresolved \hii\ Regions}
To determine the fraction of DE in the disk of NGC~4254 it is extremely 
important to verify that the DE that we detect is a distinct source of 
emission in various SFR tracer maps, and that, it is not simply the 
emission spreads out from the star-forming regions. If, for example, 
the extended tail of the PSF contains a large fraction of the source 
flux, light from star-forming \hii\ regions might scatter over a 
considerable area. This will bias the estimate of DE. 

To understand the nature of the bias we need to have clear picture about 
the shapes of the PSFs in the SFR tracer images. 
The PSF of these images show distinctive characteristics. For example, 
the PSF of \immir\ image has a complex pattern showing first and second 
Airy rings with radially extending artifacts. The linear scale of the 
second Airy ring from the center of the PSF is $\sim 20\arcsec$.  While 
approximately 85\% of the total source flux is contained within the 
central peak of the PSF which has a Gaussian shape with FWHM of 6\arcsec, 
more than 99\% flux is contained with an aperture of diameter 40\arcsec\ 
(see Table 1 of Engelbracht \etal\ 2007). The PSF of {\em GALEX} NUV 
channel varies along the field-of-view from a symmetric 2D Gaussian 
profile to extended structure further from the center. 
The FWHM of the PSF contains more than 80\% of the source flux and 95\% 
of the total flux is contained within an aperture size of 40\arcsec\ in 
diameter (see Fig. 12 of Morrisset \etal\ 2007).  
The FWHM of \imhal\ map is fairly consistent with a 2D symmetric 
Gaussian. For this map, we find two isolated field stars far from the 
source and construct their light profiles. We find that 100\% of the 
stellar light is confined within a region of radius 15\arcsec.

The smallest filter scale used in our study is 75\arcsec. By construction, 
the median filter will remove any sub-structure of the map whose linear 
dimension is equal to or smaller than half of the filter scale, i.e. 
38\arcsec\ in this case. For larger filter scales sub-structures of even 
more larger dimension will be removed. This implies that for MIR 
24 \um\ map, at 75\arcsec\ filter scale, the estimate of DE would 
have $\sim 15\%$ of the total emission that would come from the 
star-forming regions and not from the truly diffuse component un-associated 
with the region. For {\em GALEX} map one would expect a similar amount 
of contribution from the star-forming regions. For optical map it is 
insignificant. Note that with progressively larger filter scale the 
contribution of scattered light from \hii\ regions becomes negligible.

Studies of M~31 and Magellanic Clouds and Local group show that much of 
the diffuse materials are resolved into bubbles, filaments, loops, and 
shell like structures extending from few hundreds pc up to 1 kpc in the 
disk. These structures not only surround OB associations but also show 
no associations with stellar components. (Walterbos \& Braun 1994; 
Kennicutt \etal\ 1995). 
In galaxies beyond Local groups the extent of most of these features 
along with faint un-resolved \hii\ regions will fall below the 
resolution and hence will smear out. 
Therefore, any emission outside the bright, resolved \hii\ regions is 
classified as diffuse (Kennicutt \etal\ 1995). 
Comparing luminosity functions of \hii\ regions and DE Hoopes \etal\ 
(1996) showed that the contribution from the un-resolved \hii\ regions 
is around 5\%. 
The major contributing factor in the uncertainty of diffuse fraction 
estimation is the image PSF which, for this study, varies with the 
filter scale. For this analysis, however, we assign 15\% uncertainty 
in the estimate of diffuse fraction at all filter scales.


\end{document}